\documentclass[aps,prd,amsmath,amssymb,preprintnumbers,12pt]{revtex4}
\usepackage{graphicx}
\usepackage{color}

\newcommand{\bmp}{\noindent\begin{minipage}{16cm}}
\newcommand{\emp}{\end{minipage}\vskip 7mm} 

\usepackage{bm}
\usepackage{bbm}
\usepackage{pxfonts}

\newcommand{\be}{\begin{eqnarray}}
\newcommand{\ee}{\end{eqnarray}}

\newcommand{\beq}{\begin{eqnarray}}
\newcommand{\eeq}{\end{eqnarray}}
\newcommand{\bea}{\begin{eqnarray}}
\newcommand{\eea}{\end{eqnarray}}
\newcommand{\ba}{\begin{array}}
\newcommand{\ea}{\end{array}}
\newcommand{\bi}{\begin{itemize}}
\newcommand{\ei}{\end{itemize}}
\newcommand{\bn}{\begin{enumerate}}
\newcommand{\en}{\end{enumerate}}
\newcommand{\bc}{\begin{center}}
\newcommand{\ec}{\end{center}}

\newcommand{\gsim}{\lower.7ex\hbox{$\;\stackrel{\textstyle>}{\sim}\;$}}
\newcommand{\lsim}{\lower.7ex\hbox{$\;\stackrel{\textstyle<}{\sim}\;$}}

\definecolor{rossoCP3}{cmyk}{0,.88,.77,.40}

\begin{document}

\title{\Large  \color{rossoCP3}  Discovering  a Light Scalar or Pseudoscalar \\ at  \\ The Large Hadron Collider}
\author{Mads T. {Frandsen}$^{\color{rossoCP3}{\clubsuit}}$}
\email{m.frandsen1@physics.ox.ac.uk}
\author{Francesco {Sannino}$^{\color{rossoCP3}{\varheartsuit}}$}
\email{sannino@cp3.dias.sdu.dk}
 \affiliation{$^{\color{rossoCP3}{\clubsuit}}$Rudolf Peierls Centre for Theoretical Physics, University of Oxford, Oxford OX1 3NP, United Kingdom}
\affiliation{
$^{\color{rossoCP3}{\varheartsuit}}${ CP}$^{ \bf 3}${-Origins} and the Danish Institute for Advanced Study \\
University of Southern Denmark, Campusvej 55, DK-5230 Odense M, Denmark.
}


\begin{abstract}
The allowed standard model Higgs mass range has been reduced to a region between 114 and 130 GeV or above 500 GeV, at the 99\% confidence level, since the Large Hadron Collider  (LHC) program started. Furthermore some of the experiments at Tevatron and LHC observe excesses that could arise from a spin-0 particle with a mass of about 125 GeV. It is therefore timely to compare the standard model Higgs predictions against those of a more general new spin-0 state, either scalar or pseudo-scalar. Using an effective Lagrangian approach  we investigate the ability to discriminate between a scalar or pseudoscalar, stemming from several extensions of the standard model, at the LHC.  We also discuss how to use experimental results to disentangle whether the new state is elementary or composite. 
 \\
[.5cm]
{
\small \it {Preprint: CP$\,^3$-Origins-2012-005 \& DIAS-2012-06}}
\end{abstract}

\maketitle
\newpage
\section{Introduction}
Recent results from the Large Hadron Collider (LHC) experimental collaborations exclude the SM Higgs mass range between 130 GeV and 500 GeV at the 99\% confidence level \cite{:2012si,Chatrchyan:2012tx,ATLAS-CONF-2012-019,CMS-PAS-HIG-12-008,sandra,cmsmoriond} while the combined LEP2 results exclude it below 114.5 GeV at the 95\% confidence level \cite{Barate:2003sz}. At the same time both the ATLAS and CMS experiments at LHC and the CDF and D0 experiments at Tevatron \cite{FERMILAB-CONF-12-065-E,tevatronmoriond} observe excesses in certain channels around the 125 GeV region. It is therefore relevant to compare the standard model Higgs predictions against those of a more general spin-0 state, either scalar ($S$)  or pseudo-scalar ($P$), singlet with respect to the SM gauge and flavor symmetries.

We start with the SM Lagrangian without the Higgs and then add to this theory a new spin-0 state transforming as a matter field under the unbroken symmetries, using an effective Lagrangian with the operators ordered in mass dimension.  

In section~\ref{General framework} we introduce the effective Lagrangian and deduce the relevant cross-sections and partial widths, while the phenomenological setup is summarized in section~\ref{Phenomenology}.

In section~\ref{elementary} we start by investigating the case of an elementary scalar or pseudoscalar particle. In particular we consider the case in which the particle is either fermiophobic or gauge phobic,  with the pseudoscalar being naturally gauge phobic, in the investigated mass range. We also consider $b$-phobic scalar states.

In section \ref{LCS} we study signals from composite scalars and pseudscalars arising in Technicolor type extensions of the SM in which the underlying technifermions do not carry ordinary color. Several features are similar to the elementary case, 
 however, due to the presence of new technifermions the pseudoscalar decay rate into di-photons is enhanced. This fact can be used to disentangle the underlying Technicolor dynamics. 

We also point out that production of the new spin-0 state, either scalar or pseudo-scalar, in association with SM gauge bosons, can be resonantly enhanced due to the presence of new composite vector bosons \cite{Belyaev:2008yj}.  In fact, the associate production with a light composite resonance can be comparable to the SM Higgs gluon-fusion production,  leading to the interesting possibility of a di-photon signal comparable to the SM case, even for a fermiophobic spin-0 state.

For the various proposed signals we estimate the relevant significance at $5$ and $20~{\rm fb}^{-1}$ corresponding to the current and expected integrated luminosity for 2012 at  LHC . We summarize our results in section~\ref{summary}.

\section{Effective Lagrangian for spin-0 states}
\label{General framework}
The SM gauge bosons and fermions are the only experimentally observed states, until now. On theoretical grounds the internal consistency of the SM does require the presence of these states.  Neutral scalar states can be invoked to ensure perturbative unitarization of the longitudinal WW scattering amplitude. The SM Higgs provides a simple solution to this issue. However, there are well known cases where the unitarization problem can be resolved using different mechanisms \cite{Bagger:1993zf,Foadi:2008xj}. We shall therefore not be concerned with a specific unitarization model but rather introduce a new spin-0 state, either scalar or pseudoscalar, in a general way and study the associated phenomenology.

We write the electroweak and the Yukawa sectors of the SM involving only the {\it observed} particles, using a nonlinear realizations associated to the quotient  space $SU(2)_W\times U(1)_Y / U(1)_{EM}$, e.g.  \cite{Larios:1996ib}:
\begin{eqnarray}
\mathcal{L}&=&\frac{v^2}{4} {\rm Tr} [D_\mu U^\dagger D^\mu U]  -\frac{1}{2}{\rm Tr}\left[ {W}_{\mu\nu} {W}^{\mu\nu}\right]
-\frac{1}{4}{B}_{\mu\nu} {B}^{\mu\nu} +  i \bar{\psi} {\gamma_\mu D^{\mu}} \psi \nonumber \\ 
&+&  m_u \overline{\psi}_L U^\dagger \frac{1+\tau^3}{2} \psi_R +m_d \overline{\psi}_L U^\dagger \frac{1-\tau^3}{2} \psi_R     
\end{eqnarray}
Here $v \simeq 246$~GeV is the electroweak scale, $U=e^{i \frac{\pi^a \tau^a}{v}}$, ${\rm Tr}[\tau^a \tau^b]=\frac{1}{2} \delta^{ab}$,  $D_\mu U\equiv \partial_\mu U -i g W_\mu U + i g B_\mu U$ and $\psi$ denotes the SM fermions with appropriate quantum numbers. 
We now introduce a scalar ($S$) and a pseudo-scalar ($P$) state singlets under the SM gauge group, and start by classifying the various operators ordering them in mass dimension. These two states are not allowed to acquire a vacuum expectation value (VEV) since we assume they already represent the physical fluctuations around the VEV emerging from some unspecified dynamics. We denote the new scale, common to $S$ and $P$, with $\Lambda$. This new mass scale can be lighter or heavier than the scale $v$.

The Yukawa type interactions of $S$ or $P$ with the SM fermions $\psi$ and any new Dirac fermions $\Psi$ are:
\begin{align}
  \frac{m_{\psi}}{v} (1+ {y_{S\psi}} \frac{v}{\Lambda}) S \, \overline{\psi} \psi  +  i\, m_{\psi} \frac{y_{P\psi}}{\Lambda} P \,\overline{\psi}\gamma_5 \psi   + 
   \frac{m_{\Psi}}{\Lambda}  (y_{S\Psi} S \,\overline{\Psi} \Psi + i y_{P\Psi} P  \,\overline{\Psi}\gamma_5 \Psi )
  \ . 
 \label{Eq:Yukawa}
    \end{align}
Here we are considering, after electroweak symmetry breaking, each individual SM fermion separately. In the SM limit we have
\begin{eqnarray}
y_{S\psi} = y_{S\Psi} =  0 \ , \qquad {\rm and}  \qquad y_{P\psi} = y_{P \Psi} = 0 \ . 
\end{eqnarray}
For the interactions with the SM gauge fields, also linear in $S$ and $P$, and to the third and fifth order in mass dimension we have:
 \beq 
\mathcal{L}^{(3)}&=&g^2 \frac{v}{2} \left(1 +g_{SWW}^{(1)} \frac{\Lambda}{v}\right)\, S {W^+}_{\mu}{W^-}^{\mu}+ g^2 \frac{v}{4 \cos^2\theta_w} \left(1 +g_{SZZ}^{(1)} \frac{\Lambda}{v}\right)\, S {Z}_{\mu}{Z}^{\mu} 
\label{tree}
\eeq
\beq
  \mathcal{L}^{(5)}_S &=&
\frac{1}{4}{g^{(2)}_{SWW }}{\Lambda}^{-1}\,S  W^+_{\mu\nu}{W^-}^{\mu\nu}
+ \frac{1}{4}{g^{(2)}_{SZZ}}{\Lambda}^{-1}\,SZ_{\mu\nu} Z^{\mu\nu}
+ \frac{1}{4}{g^{(2)}_{S\gamma\gamma}}{\Lambda}^{-1}\ {S}\ F_{\mu \nu} F^{\mu \nu} \nonumber \\&&+ \frac{1}{4}g^{(2)}_{SZ\gamma} {\Lambda}^{-1}{P}\ Z_{\mu \nu} F^{\mu \nu}+\frac{1}{4}g^{(2)}_{S gg}{\Lambda^{-1}}{S}\sum_{a=1}^8 G_{\mu \nu}^a G^{a\mu \nu}\\ 
  \mathcal{L}^{(5)}_P
&=&
\frac{1}{4}{g_{PWW}}{\Lambda}^{-1}\,P {\widetilde W}^+_{\mu\nu}{W^-}^{\mu\nu}
+ \frac{1}{4}{g_{PZZ}}{\Lambda}^{-1}\,PZ_{\mu\nu}\widetilde{Z}^{\mu\nu}
+ \frac{1}{4}{g_{P\gamma\gamma}}{\Lambda}^{-1}{P}\ F_{\mu \nu} \widetilde{F}^{\mu \nu} \nonumber \\&&+ \frac{1}{4}g_{PZ\gamma} {\Lambda}^{-1}{P}\ Z_{\mu \nu} \widetilde{F}^{\mu \nu}+\frac{1}{4}g_{P gg}{\Lambda^{-1}}{P}\sum_{a=1}^8 \widetilde{G}_{\mu \nu}^a G^{a\mu \nu}\ \ , 
\eeq
where $F^{\mu\nu}, G^{\mu\nu}$ are the field strength tensors of the photon and the gluon respectivevly, we have defined $\widetilde{F}_{\mu\nu}=\epsilon_{\mu\nu\rho\sigma} F^{\rho\sigma}$ and the tree-level SM is recovered when: 
\begin{equation}
g_{SWW}^{(1)} = 0 \ , \qquad {\rm and} \qquad g_{SZZ}^{(1)} = 0 \ . \end{equation} 
 If the underlying model is known one can predict the coefficients of  all the terms. If the specific model is not known the effective low energy theory has a large number of unknown couplings, and to be more predictive, we assume that the dimension five operators are determined, at the one-loop level, using \eqref{Eq:Yukawa} and \eqref{tree}.

 The $S$ and $P$ di-photon and di-gluon partial widths are  compactly written in terms of the coefficients of the relevant dimension five operators and  read \cite{Spira:1995rr}:
 \bea
\Gamma (S\to gg)
&=& \frac{1}{8\pi} \, \vert g^{(2)}_{S gg} \vert^2
\frac{m_S^3}{\Lambda^2} \ , \qquad
\Gamma (P\to gg)
= \frac{1}{2\pi} \, \vert g_{P gg} \vert^2
\frac{m_P^3}{\Lambda^2} 
\ ,\\
\label{eq:sgaga}
 \Gamma (S\to \gamma\gamma)
&=& \frac{1}{64\pi} \, \vert {g}_{S\gamma\gamma}^{(2)} \vert^2\,
\frac{m_S^3}{\Lambda^2} \ , \qquad
\Gamma (P\to \gamma\gamma)
= \frac{1}{16\pi} \, \vert {g}_{P\gamma\gamma} \vert^2\,
\frac{m_P^3}{\Lambda^2} \ , 
 \eea
We can express the coefficients in terms of couplings of the lower dimensional operators as follows:
\begin{eqnarray}
g_{S\gamma\gamma}^2 &=&
\frac{\alpha \Lambda}{2 v\pi}
\left| \sum_\psi  N_c e_\psi^2 g_{S\psi} A_\psi^{ S} + \sum_{\Psi}  d(R_\Psi) e_\Psi^2 g_{S\Psi} A^S_{\Psi}+
g_{S V} A_V^{ S} \right|
\\
g_{P\gamma\gamma}^2 &=&
\frac{\alpha \Lambda}{2 v\pi}
\left| \sum_\psi N_c e_\psi^2 g_{P\psi} A^P_\psi + \sum_{\Psi}  d(R_\Psi) e_\Psi^2 g_{P\Psi} A^P_{\Psi}
\right| \nonumber \\
g_{S gg}  &=&  \frac{\alpha_s \Lambda}{4v \pi}
\left| \frac{3}{4} \sum_{q\in \psi}g_{S q} A^S_q  + \sum_{Q\in \Psi} g_{S Q} A^S_Q \right|  \ ,
\\ 
g_{P gg}  &=&  \frac{\alpha_s \Lambda}{4v \pi}
\left| \sum_q g_{P q} A^P_q +  \sum_Q g_{PQ} A^P_Q\right|
\label{loopcouplings}
\end{eqnarray}where $q,Q$ denote colored SM quarks and new  colored fermions respectively while $d(R_\Psi)$ is the dimension of the representation of the new fermions $\Psi$ with respect to QCD and any new gauge groups.
 
 The spin 1, spin 1/2 and spin 0 amplitudes read to lowest
order for the scalar S \cite{Djouadi:2005gi}
\begin{eqnarray}
A_V^S     & = & - [2\tau_V^2+3\tau_V+3 (2\tau_V-1)f(\tau_V)]/ \tau_V^2 \to -7 \ , \quad {\rm with} \quad  \tau_V \to 0 \nonumber \\
A_{\psi}^S & = & 2 [\tau_\psi +(\tau-1)f(\tau_\psi)]/\tau_\psi^2 \to \frac{4}{3} \ ,  \quad {\rm with} \quad  \tau_\psi \to 0 \ ,
\label{AAA}
\end{eqnarray}
and for the pseudoscalar P
\begin{eqnarray}
A_{\psi}^P & = & f(\tau_\psi)/\tau_\psi \to 1 \ ,   \quad {\rm with} \quad  \tau_\psi \to 0 \ , 
\end{eqnarray}
where  $\tau_{V, \psi, S} = m_\Phi^2/4 m_{V, \psi, S}^2$ and the function $f(\tau_{V, \psi, S})$ is given by
\begin{eqnarray}
f(\tau)=\left\{
\begin{array}{ll}  \displaystyle
\arcsin^2\sqrt{\tau} & \tau\leq 1 \\
\displaystyle -\frac{1}{4}\left[ \log\frac{1+\sqrt{1-\tau^{-1}}}
{1-\sqrt{1-\tau^{-1}}}-i\pi \right]^2 \hspace{0.5cm} & \tau>1
\end{array} \right.
\label{eq:ftau}
\end{eqnarray}
In these formulae $\Phi$ is either one of the scalars and $m_{V, \psi, S}$ are the masses of the states, spin-1, spin-1/2 or spin-0, running in the loop. For new strongly interacting fermions, such as the techniquarks in Technicolor, the dynamical mass is intrinsically linked to the decay constant $F_\Phi$ via the Pagels-Stokar formula $M_{\Psi} \sim {2\pi F_\Phi}/{\sqrt{d(R_\Psi)}}$. Finally $g_{\Phi (V, \psi, S)}$ is the coupling of the (pseudo)scalar field to the particles running in the loop.
These couplings, following from the effective Lagrangian, are reported in table~\ref{loopcouplingsI} for the SM Higgs (H) as well as $S$ and  $P$. 
\begin{table}[hbt]
\renewcommand{\arraystretch}{1.5}
\begin{center}
\begin{tabular}{|lc||ccc|} \hline
\multicolumn{2}{|c||}{~~$\Phi$~~} 
& $g_{\Phi \psi}$ &  $g_{\Phi W}$ &  $g_{\Phi \Psi}$ \\ \hline \hline
& $H$~~ & $1$ & 1 & 0  \\ \hline
 & $S$ ~~& $1+\frac{v}{\Lambda}  y_{S \psi} $~~  &
$1+g^{(1)}_{S WW}  \frac{v}{\Lambda}$  ~~& $ \frac{v}{\Lambda}  y_{S \Psi}$    \\ \hline
& $P$ ~~& $ \frac{v}{\Lambda}  y_{P \psi} $  &
$0$  & $  \frac{v}{\Lambda}  y_{P \Psi} $   \\ \hline
\end{tabular}
\renewcommand{\arraystretch}{1.2}
\caption{
{\small Couplings of the spin-0 state $\Phi$ to SM fermions $\psi$, $W$ bosons, and any new fermions beyond the SM $\Psi$, entering the formulas for loop induced decays into massless gauge bosons in Eq.~(\ref{loopcouplings}).
$H$ denotes the SM Higgs, $S$ is a generic scalar and $P$ a generic pseudoscalar}}
\end{center}
\label{loopcouplingsI}
\end{table}

\section{Phenomenological Set Up}
\label{Phenomenology}
It is phenomenologically convenient to broadly classify spin-0 states, compared to the SM Higgs, as fundamental vs composite and fermiophobic vs gaugephobic. Furthermore a generic scalar can be CP-even or CP-odd. 

For the generic spin-0 state  $\Phi=\{S , P\}$ we will investigate the four production mechanisms corresponding to the main production mechanisms for the SM Higgs boson, i.e. gluon and vector-boson fusion, production in association with a SM $W$ or $Z$ boson and $t \bar{t}$ fusion:
\beq
\sigma(pp\to \Phi ) \ , \quad \sigma(pp\to \Phi qq') 
\ , \quad
\sigma(pp\to \Phi \, V) \ , \quad 
\quad \sigma(pp\to \Phi t\bar{t}) \ , 
\eeq
where $V=W, Z$ and $q,q'$ denote SM quarks. 

We are interested in the enhancement ratios of these cross-sections at LHC for a scalar $\Phi$ relative to the SM Higgs and therefore define production enhancement factors $\kappa$ as 
\beq
\label{Eq:kappaprod}
\kappa_{\Phi }^{\rm prod} &\equiv& \frac{\sigma(pp\to \Phi )}{\sigma(pp\to H )}\ , \quad \kappa_{\Phi V}^{\rm prod} \equiv \frac{\sigma(pp\to \Phi V)}{\sigma(pp\to H V)} 
\  , \\ 
\kappa_{\Phi t\bar{t} }^{\rm prod}  &\equiv& \frac{\sigma(pp \to  \Phi t\bar{t}) }{\sigma(pp \to  H\ t\bar{t})}    \ , \quad 
\kappa_{\Phi qq'}^{\rm prod} \equiv \frac{\sigma(pp \to \Phi qq')}{\sigma(pp \to H qq'))}  \ . 
\eeq
For the SM Higgs, the dominant initial states in gluon fusion and top fusion production are gluons while in associate and vector boson fusion production they are quarks. 
We give the LHC production cross-sections at 7 and 8 TeV from \cite{ATLAS:1303604,cs} of a SM Higgs in these channels in table~\ref{table:smcs}.
\begin{table}[!h]
\caption{SM Higgs production cross-sections in pb for $m_H=126$ GeV from \cite{ATLAS:1303604,cs}.}
\vspace*{-6mm}
\label{table:smcs}
\vspace{3mm}
\footnotesize
\begin{center}
\begin{tabular}{|l|r|r|r|r|r|}
\hline
$\sqrt{s}$ [TeV] & $\sigma(pp\to H )$  & $\sigma(pp \to H qq')$& $\sigma(pp\to H W)$ & $\sigma(pp\to H Z)$& $\sigma(pp \to  H\ t\bar{t})$
\\\hline\hline
7 & $15$ & $1.2 $ & 0.56 & 0.31 & 0.084
\\\hline\hline
8 & 19 & $1.45$ & $0.69 $ &  0.36 & na
  \\\hline
\end{tabular}\end{center}
\end{table}
The equivalent enhancement factors of the $\Phi$ decay modes are
\be
\label{Eq:kappaprod}
\kappa_{\Phi, V_1 V_2 }^{\rm dec} \equiv \frac{\Gamma (\Phi \to V_1 V_2)}{\Gamma (H \to V_1 V_2)}\ , \quad \kappa_{\Phi, \psi\bar{\psi} }^{\rm dec} \equiv \frac{\Gamma (\Phi \to \psi\bar{\psi})}{\Gamma (H \to \psi\bar{\psi})}\ , 
\eeq
where $\psi$, as above, is any SM fermion and $V_i=W,Z,\gamma, g$. When both are massive gauge bosons, one of them will be off-shell for $m_\Phi \sim 126$ GeV. Finally we define the total enhancement factors $R\equiv \sigma/\sigma_{\rm SM}$ which in the narrow width approximation, and choosing the di-photon channel as example, can be written as 
\beq
\label{Rgammagamma}
R_{\gamma\gamma}^\Phi =\frac{\sigma(pp\to \Phi ) }{\sigma(pp\to H)} \frac{{\rm BR}[\Phi \to \gamma\gamma]}{ {\rm BR}[H\to \gamma\gamma]}=
\kappa_{\Phi }^{\rm prod}  \kappa_{\Phi, \gamma \gamma }^{\rm dec}  \frac{\Gamma[H]}{\Gamma[\Phi]} \ .
\eeq
Note that the subscript on $R$ only refers to the final state. The total enhancement factors we consider in this study are \footnote{Note that in the subscripts of the enhancement factors $R$ we use $ttbb$ for $t\bar{t} b\bar{b}$ and $bb$ for $b\bar{b}$.}
\beq
R^\Phi_{V_1 V_1} \ ; \quad R^\Phi_{ttbb}  \ ;  \quad R^\Phi_{V_1 V_2 q q'}  \ ; \quad  R^\Phi_{V_1 V_2 \  V} \ , \ R^\Phi_{bb V}
\eeq
$R^\Phi_{V_1 V_1}$ arises from gluon fusion production of $\Phi$ which decays into vectors, $R^\Phi_{ttbb}$ from the top-fusion process with $\Phi$ decaying into $b$-quarks,  $R^\Phi_{V_1 V_2 q q'}$ from vector boson fusion with $\Phi$ decaying into vectors and finally $R^\Phi_{V_1 V_2\ V},R^\Phi_{bb V} $ arise from associate production of $\Phi$ which subsequently decays into vectors or $b$-quarks.

We define the expected significance of $\Phi$ in a given final state $X$ by ${\cal S}^\Phi_{X}$.  In terms of the signal and background cross-sections $\sigma_X^\Phi$, $\sigma_{\rm  bg}$ and the luminosity $L$ this is given by 
\beq
{\cal S}_{X}^\Phi(L) \equiv \frac{\sigma_X^\Phi }{\sqrt{\sigma_{\rm bg}}} \sqrt{L} =  {\cal S}^{H}_X (L) R_{X}^\Phi \ , 
\eeq
where, in the last equality, we have expressed the expected significance in terms of that for the SM Higgs boson $H$ and the total enhancement factor $R_{X}^\Phi$.

Estimates for the expected significance for a SM Higgs signal are given in table~\ref{table:smhiggssig}. These estimates assume that the significance simply scales as $\sqrt{L}$ and are based on significances quoted for $\sim 5 \, {\rm fb}^{-1}$ \cite{:2012si,ATLAS-CONF-2012-019}. 
We refer to the appendix~\ref{Significance} for a discussion of the numbers quoted for $S^H_{ttbb}$.
\begin{table}[!h]
\caption{Estimates of the expected significance, in numbers of $\sigma$, of a SM Higgs signal at $m_H=126$ GeV and $\sqrt{s}=7$ TeV based on \cite{:2012si,ATLAS-CONF-2012-019,Plehn:2009rk}}
\vspace*{-6mm}
\label{table:smhiggssig}
\vspace{3mm}
\footnotesize
\begin{center}
\begin{tabular}{|l|r|r|r|r|}
\hline
 & $S^H_{\gamma\gamma}$  & $S^H_{ZZ^*}$& $S^H_{WW^*}$ & $S^H_{ttbb}$
\\\hline\hline
ATLAS,  7 TeV  & $0.64 \sqrt{L}$ \cite{:2012si}& $0.64 \sqrt{L} $  \cite{:2012si} & $0.74 \sqrt{L}$ \cite{ATLAS-CONF-2012-019}& $0.06 \sqrt{L}$ \cite{Plehn:2009rk}  
\\\hline\hline
\end{tabular}\end{center}
\label{tableofsig}
\end{table}

We compute the branching ratios and the width of the SM Higgs at one-loop level for the $\gamma\gamma$ and $gg$ decay modes and at tree-level for all other decay modes. This level of precision is sufficient for our discussion as we are interested in the scaling of the above cross-section ratios in simple scenarios. The results are given below in table~\ref{table:smbr} for $m_H=126$ GeV.
\begin{table}[!h]
\caption{Branching ratios and total width of the SM Higgs boson computed at leading order for $m_H=126$ GeV.}
\vspace*{-6mm}
\label{table:smbr}
\vspace{3mm}
\footnotesize
\begin{center}
\begin{tabular}{|l|r|r|r|r|r|r|r|r|}
\hline
 BR($H\to gg$)& BR($H\to \gamma\gamma$) & BR($H\to ZZ^*$) & BR($H\to WW^*$) &
BR($H\to b\bar{b}$) &
BR($H\to \tau\bar{\tau}$)&
BR($H\to c\bar{c}$)& $\Gamma_{tot}$ [GeV]
\\\hline\hline
 $5.5 \cdot 10^{-2}$ & $2.4 \cdot 10^{-3}$ & 0.026 & 0.23 & 0.60 & 0.06 &0.024 & $4.2 \cdot
10^{-3}$\\\hline
\end{tabular}\end{center}
\end{table}

The results of current LHC data analysis in selected channels are summarized in table~\ref{table:higgssigmoriond}. 
The local significance of the observed excesses (those expected for a SM Higgs are given in parentheses) at ATLAS  are $2.9\sigma$ ($1.4 \sigma$) in the $pp \to \gamma\gamma$ channel \cite{Collaboration:2012sk},  $2.1 \sigma$ ($1.4 \sigma$)  in the $pp \to Z Z^* \to l^+ l^- l'^+ l'^-$ channel \cite{:2012sm}, and $0.2\sigma$ ($1.6 \sigma$) in the $pp \rightarrow  W \, W^* \rightarrow \ell^+ \, \nu  \, \ell'^- \, \bar{\nu'} $ channel \cite{ATLAS-CONF-2012-019} \footnote{Note that the combined analysis \cite{ATLAS-CONF-2012-019} quotes an expected significance of  $1.6 \sigma$ in this channel with 4.7 ${\rm fb}^{-1}$ while the previous combined analysis \cite{:2012si} quoted an expected significance of 1.4$\sigma$ with 2.05 ${\rm fb}^{-1}$. Obviously in this case the expected significance does not increase as rapidly with luminosity as $\sqrt{L}$.}. For CMS they are $2.9\sigma$ ($1.4 \sigma$) in the $pp \to \gamma\gamma$ channel \cite{CMS-PAS-HIG-12-001} at 125 GeV while they are $2.7 \sigma$ at 119.5 GeV and only $1.5 \sigma$ at 126 GeV in the $pp \to Z Z^* \to l^+ l^- l'^+ l'^-$ channel \cite{Chatrchyan:2012dg}. The local significance of the observed excess at CDF and D0 is $2.9\sigma$ in the $b \bar{b}$ channel.
\begin{table} 
\setlength{\tabcolsep}{4pt}
\begin{tabular}{||c|c|c|c||} 
\hline \hline 
Channel [Exp] &  L  $[{\rm fb}^{-1}]$ & $m_H [{\rm GeV}]$  & $R$ fit ($R$ limit)  
\\
\hline
$pp \rightarrow \gamma \, \gamma \,\, [{\rm ATLAS}]$  & 4.9 & $126.5 \pm 0.7  $ \cite{Collaboration:2012sk} & $2^{+0.9}_{-0.7}$  (2.6)      \cite{ATLAS-CONF-2011-163}.
\\
$pp \rightarrow  \gamma \, \gamma  \, [{\rm CMS}]$  & 4.8& $125 $ & $1.65^{0.67}_{-0.6} (2.9)$ \cite{CMS-PAS-HIG-12-001}
\\
$pp \rightarrow  W \, W^* \rightarrow \ell^+ \, \nu  \, \ell'^- \, \bar{\nu'} \,\, [{\rm ATLAS}] $ & 4.7&  no excess & $0.16^{+0.6}_{-0.6}$ (1.3) @ 126 GeV\cite{ATLAS-CONF-2012-012}    
\\
$pp \rightarrow  W \, W^* \rightarrow \ell^+ \, \nu  \, \ell^- \, \bar{\nu} \,\, [{\rm CMS}] $ & 4.6& no excess & $0.4^{+0.6}_{-0.55}$ \cite{CMS-PAS-HIG-12-008} (1.35)  \cite{Chatrchyan:2012ty} @ 125 GeV 
  \\
$pp \rightarrow  Z \, Z^* \rightarrow \ell^+ \, \ell^- \, \ell'^+ \, \ell'^- \,\, [{\rm ATLAS}] $ & 4.8 & $125 $  \cite{:2012sm} & $1.2^{+1.2}_{-0.8}$ (4.9)  @ 126 GeV  \cite{ATLAS-CONF-2011-163}
\\
$pp \rightarrow  Z \, Z^* \rightarrow \ell^+ \, \ell^- \, \ell^+ \, \ell^- \,\, [{\rm CMS}] $ & 4.7&$125 $ & $0.58^{+1.0}_{-0.58}$   \cite{CMS-PAS-HIG-12-008} (2.5)    \cite{Chatrchyan:2012dg}
 \\
$ H \to b\bar{b} \quad [{\rm CDF/D0 \, combined}] $ & 10/9.7 &115-135 &   $1.6^{+0.6}_{-0.6}$ @ 120 GeV  \cite{FERMILAB-CONF-12-065-E,tevatronmoriond}
 \\
\hline \hline
\end{tabular}
\caption{A summary of selected Higgs search final states, including the mass range consistent with the observed excess, the best fit cross-section in terms of $R\equiv \sigma/\sigma_{\rm SM}$ where $\sigma_{\rm SM}$ is the SM Higgs cross-section. Finally the limit quoted on $R$ given in parenthesis is the $95\%$ CL. }
\label{table:higgssigmoriond} \vspace{-0.35cm}
\end{table}
A number of recent studies perform fits to the corresponding data before the updates \cite{ATLAS-CONF-2012-019,CMS-PAS-HIG-12-008,sandra,cmsmoriond} in terms of a scalar state $S$, within an effective Lagrangian framework \cite{Carmi:2012yp,Azatov:2012bz,Espinosa:2012ir}.


\section{Light Elementary Spin-0 States and Their Signatures}
\label{elementary}
It is instructive to first analyze the experimental signatures coming from light, $m_\Phi\sim 126$ GeV, elementary spin-0 states at the LHC.
In particular we will consider several relevant limits of the couplings of a SM-like Higgs scalar, as well as a pseudoscalar state.  

\subsection{Scalar with reduced $b$-Yukawa}
\label{reducedb}
A minimal way of enhancing the di-photon rate of a scalar $S$ is to reduce the $b$-Yukawa relative to the SM Higgs. 
In the limit where we turn the $S$ coupling to b-quarks to zero we have from table~\ref{table:smbr} that $\frac{\Gamma[H]}{\Gamma[S]}=\frac{1}{1-{\rm Br}[H\to \bar{b}b]}\sim 2.5$. 

\begin{figure}[htp!]
\par
\begin{center}
\includegraphics[width=7.0cm]{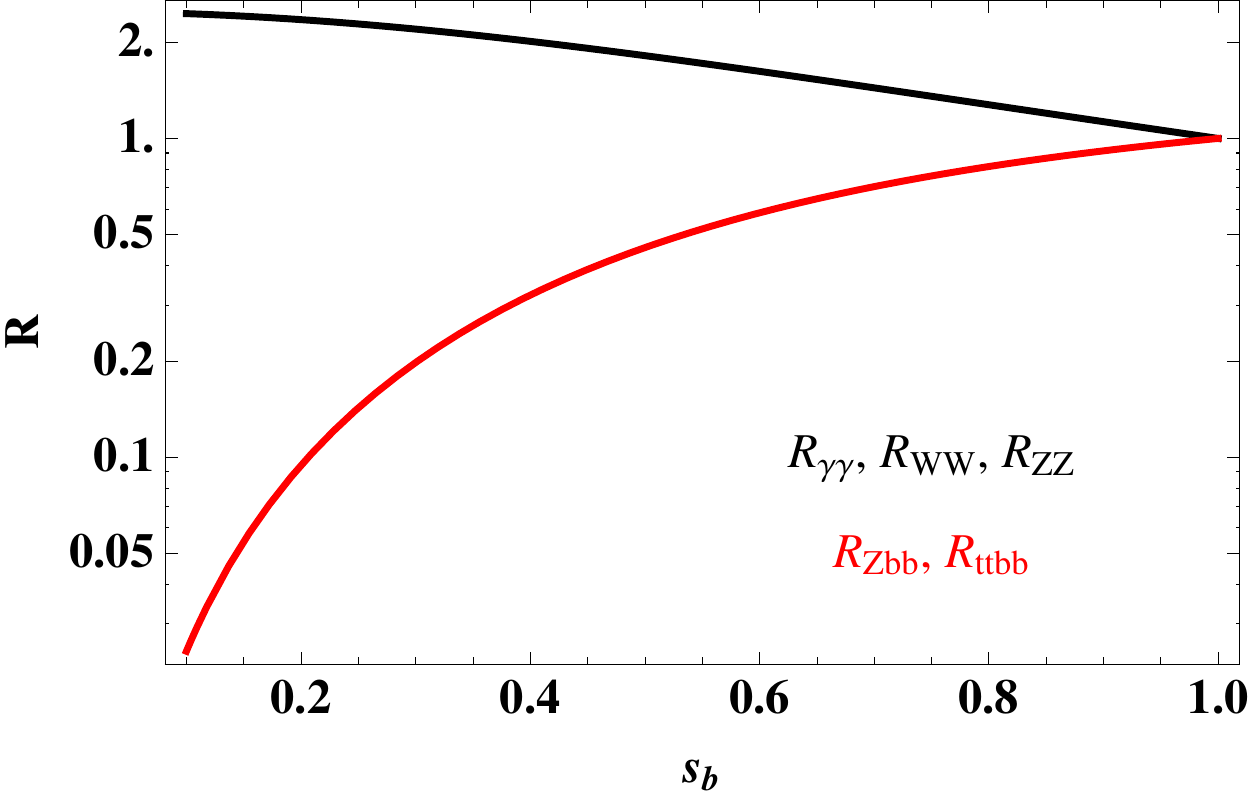}
\includegraphics[width=7.0cm]{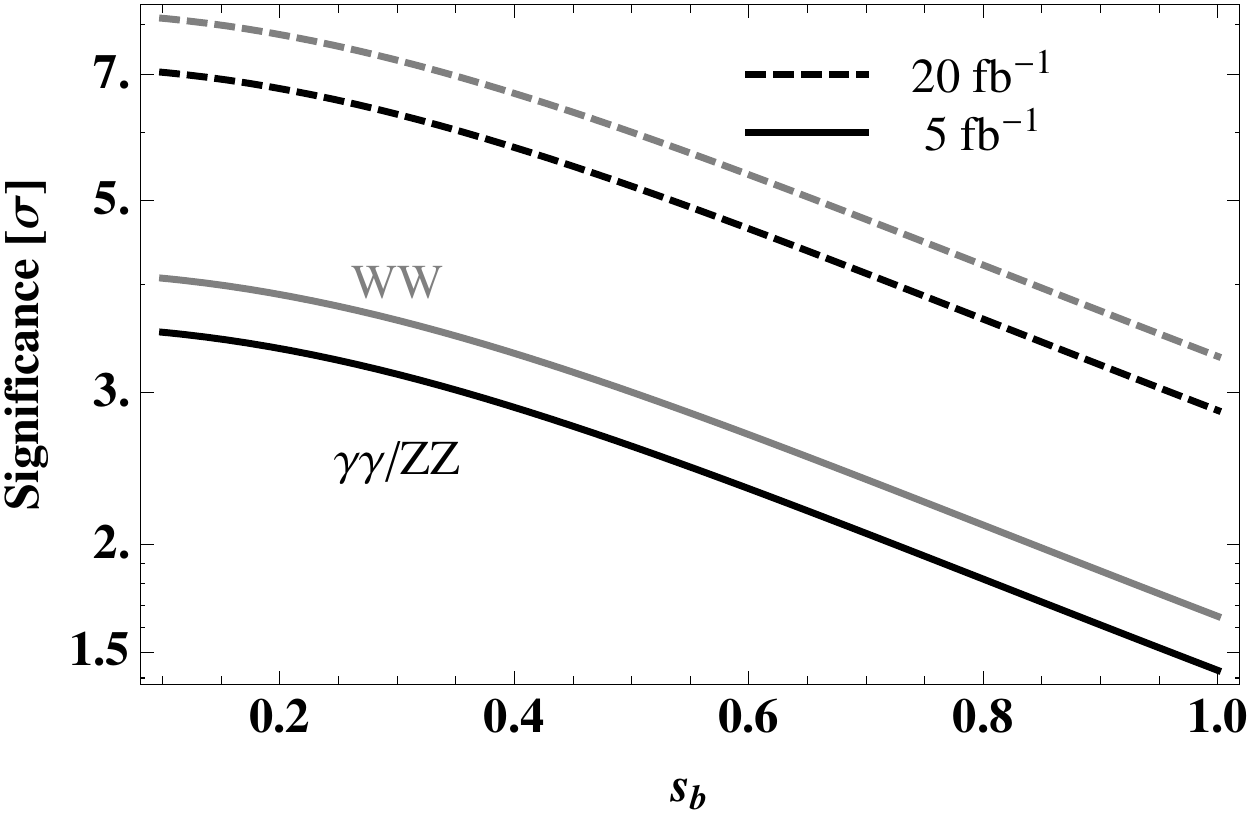}
\end{center}
\caption{Left: The enhancement ratios $R_{\gamma\gamma}=R_{WW}=R_{ZZ}$ (black) and $R_{bbZ}=R_{ttbb}$ (red) for the scalar $S$ as a function of $s_b$, the fraction of the SM $b$-Yukawa defined in section~\ref{reducedb}. 
Right: The corresponding expected significance ${\cal S}_{\gamma\gamma}^S={\cal S}_{ZZ}^S$ (black) and ${\cal S}_{WW}^S$ (gray) with 5 (solid) and 20 (dashed) ${\rm fb}^{-1}$ of data. }
\par
\label{Eq:bscaling}
\end{figure}

More generally we write the $b$-Yukawa given in Eq.~(\ref{Eq:Yukawa}) as $1+ y_{Sb}\frac{v}{\Lambda}\equiv s_b$. 
For $s_b<1$, the pattern of simultaneous enhancements of the vector decay modes and the reduction of $b\bar{b}$ final states, is given by
\beq
R_{\gamma\gamma/WW/ZZ}\sim\frac{1}{0.6 s_b^{2}+0.4} \ , \quad R_{ttbb}=R_{bb \ V} \sim  \frac{s_b^2}{0.6 s_b^{2}+0.4}  \ ,
\eeq
We show these enhancement factors in the left panel of Fig.~\ref{Eq:bscaling}.
The expected significances for the $\gamma\gamma$ and the leptonic $ZZ^*$ and $WW^*$ final states in this scenario, based on the numbers presented and discussed in table~\ref{tableofsig}, are then
\beq
{\cal S}_{\gamma\gamma}^S(L) \sim {\cal S}_{ZZ}^S(L)  \sim  0.64 \, \sqrt{L} \, R_{\gamma\gamma}^S \ , \ {\cal S}_{WW}^S(L)  \sim  0.74 \, \sqrt{L} \, R_{\gamma\gamma}^S \nonumber \ ,
\eeq
as shown in the right panel of Fig.~\ref{Eq:bscaling}.

For $R_{\gamma\gamma}^S > 1$, as the current di-photon data hints at, we expect sizeable correlated excesses in both the $WW$ and $ZZ$ channels by the end of 2012. 
If this is observed, the scenario can be further tested by the absence or strong reduction, compared to the SM Higgs, of the $b\bar{b}$ signals in the $bb V$ \cite{Butterworth:2008iy,ATL-PHYS-PUB-2009-088} and the $ttbb$ final states \cite{Plehn:2009rk}. However, we note that $R_{\gamma\gamma}^S \gtrsim 1.5$ is at odds with both ATLAS and CMS limits on $R_{WW}$ at the 95 $\%$ CL.
Model examples with a reduced $b$-Yukawa include the top version of the  {\it Private Higgs} of \cite{Porto:2007ed}, which couples to the SM top and vector bosons only. Another is the Higgs of the NMSSM where enhancements of the di-photon rate were considered in e.g. \cite{Hall:2011aa,Ellwanger:2011aa,King:2012is}.

\subsection{Complete fermiophobic Scalar}
The limit of a fully fermiophobic scalar was recently considered in \cite{Gabrielli:2012yz}. At the lowest order in the SM couplings the only production mechanisms are vector boson fusion and Higgs-strahlung. If we first assume that the vector boson couplings are those of the SM Higgs we then have
$ \frac{\Gamma[H]}{\Gamma[S]} \sim 3.8$, $\kappa_{S, \gamma\gamma }^{\rm dec}\sim 1.6$
Thus we end up with enhancements 
\beq
R_{V\gamma\gamma}=R_{q q'\gamma\gamma}\sim 6.3
\eeq
From table~\ref{table:smcs} it follows that this is sufficient for the Higgs-strahlung and VBF processes to generate a scalar production cross-section of roughly 0.8 times the size of the SM Higgs cross-section from gluon fusion at 126 GeV. This was recently observed in \cite{Gabrielli:2012yz}.
 Once the Higgs mass is below 124 GeV the cross-section is as large as the corresponding SM Higgs cross-section from gluon fusion.

A {\it purely} Technicolor extension of the SM, featuring a light composite Higgs \cite{Sannino:2004qp, Dietrich:2005jn,Dietrich:2006cm}  (sometimes also called the dilaton \cite{Matsuzaki:2012gd}) provides such a fermiophobic state.  This is so since new dynamics beyond Technicolor is needed to endow the SM fermions with mass \cite{Eichten:1979ah,Dimopoulos:1979es}. Although in a purely Technicolor extension the composite technistates do not couple directly to the SM fermions these couplings will emerge at higher orders in the SM couplings. 
However, if the fermiophobic scalar arises from new strong dynamics the di-photon decay rate is modified by the effects of technifermions which must be taken into account. We discuss this later.

\subsection{Gauge phobic Scalar}
\label{Vector phobic Scalar}
Another relevant limit is the gauge phobic scalar $S$ which couples only via Yukawas to the SM fermions. One model where such a situation is realized is the {\it Private Higgs} model  \cite{Porto:2007ed}. A gauge phobic scalar coupled to $t$ (or $b$) can provide large di-photon cross-sections. 

\begin{figure}[htp!]
\par
\begin{center}
\includegraphics[width=7.0cm]{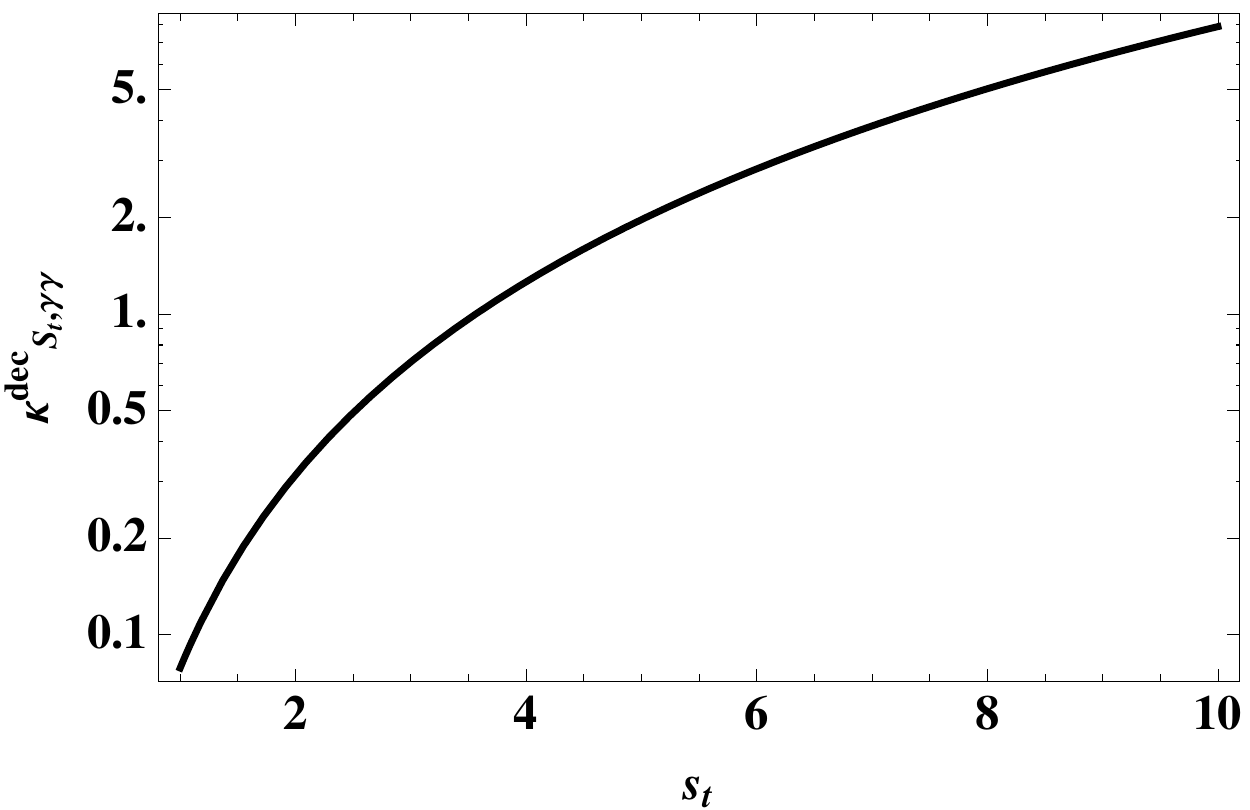}
\includegraphics[width=7.0cm]{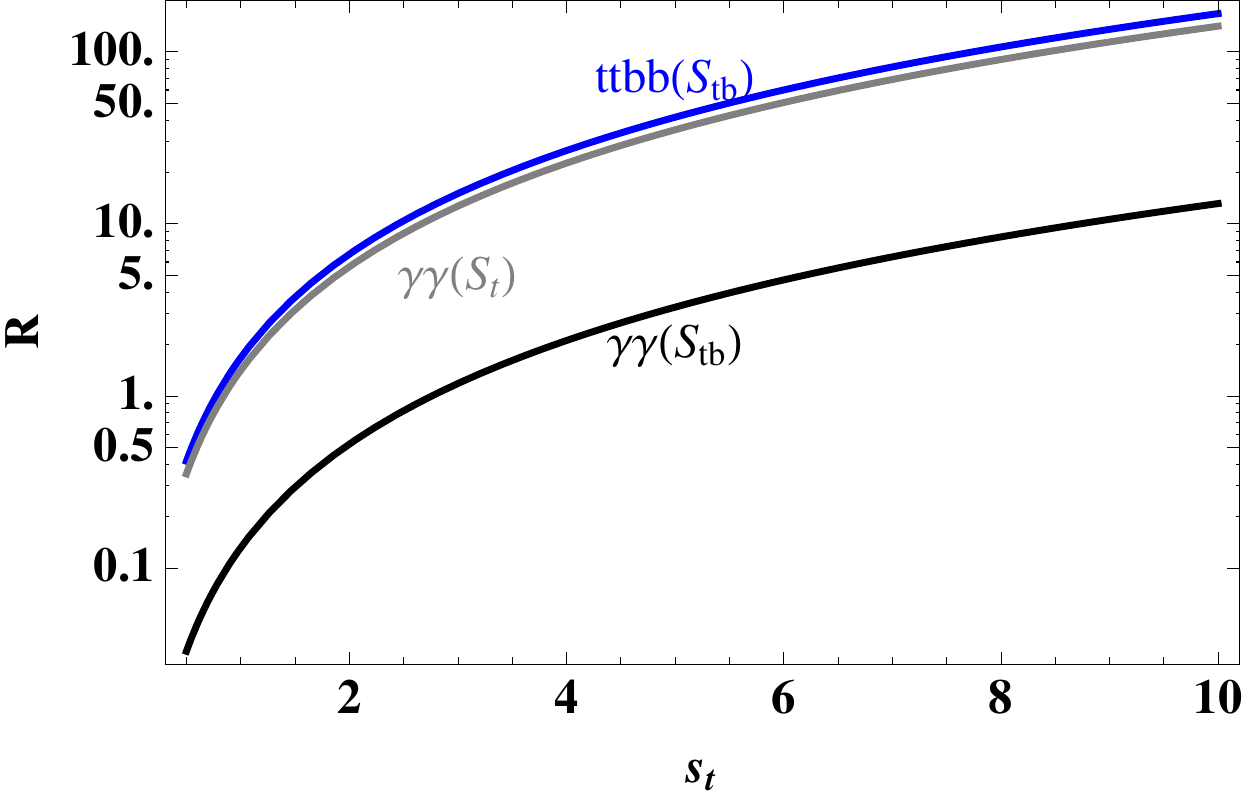}
\includegraphics[width=8.0cm]{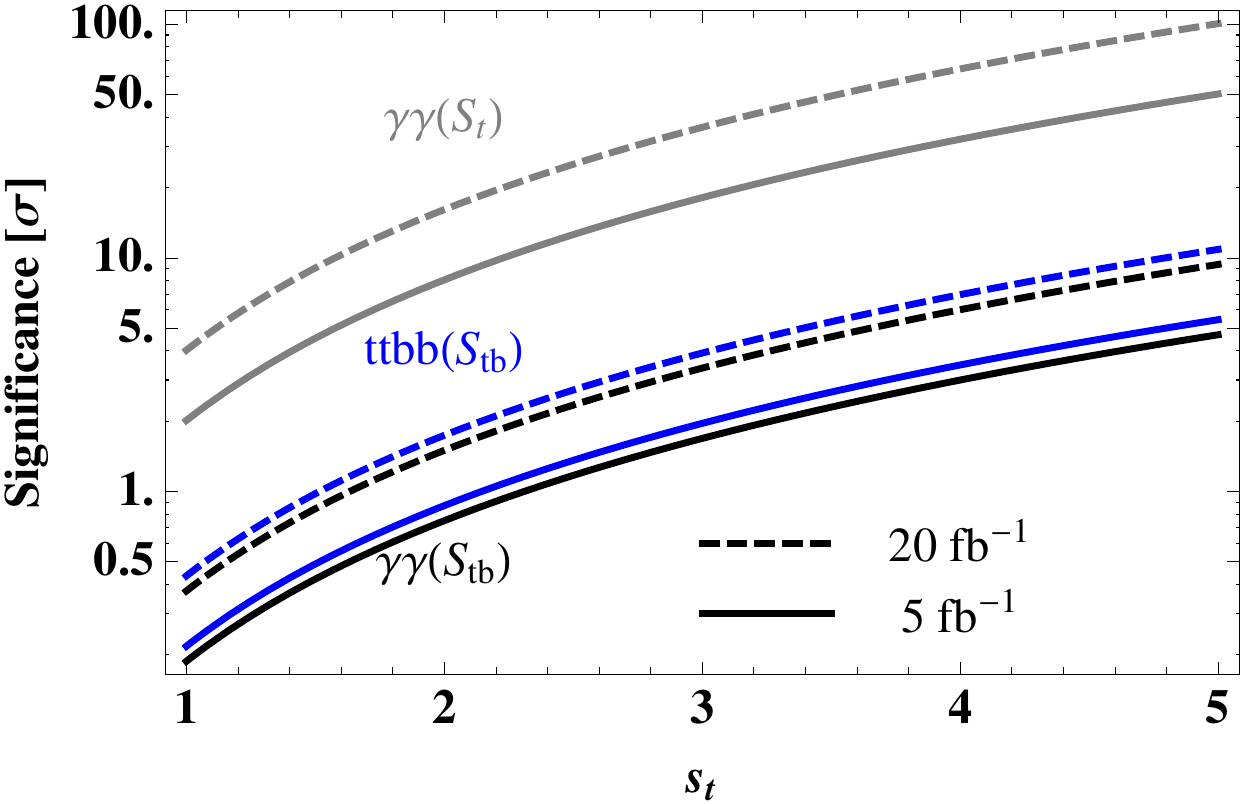}
\end{center}
\caption{Top left: The enhancement  $\kappa^{dec}_{S_t, \gamma\gamma}\simeq \kappa^{dec}_{S_{tb}, \gamma\gamma} $ of the decay width of $S_t$ and $S_{tb}$ into di-photons as a function of $s_t$, the fraction of the SM $t$-Yukawa, defined in section~\ref{Vector phobic Scalar}. 
Top right:
The corresponding total enhancement ratios $R_{\gamma\gamma}$ of $S_t$ (black dashed) and $S_{tb}$ (black solid) as well $R_{ttbb}$ for $S_{tb}$ (blue, solid) as a function of $s_t$. Bottom: The expected significance in the $\gamma\gamma$ and $ttbb$ channels as a function of $s_t$ with 5 (solid) and 20 (dashed) ${\rm fb}^{-1}$ of data.}
\par
\label{Fig:bscaling}
\end{figure}

We parameterize the top Yukawa coupling by setting $1+ y_{St}\frac{v}{\Lambda}\equiv s_t$
and distinguish between a scalar $S_t$ coupling purely via this top-Yukawa or a scalar $S_{tb}$ which couples to both the $t$ and $b$ quarks. For simplicity we first parameterize the latter case by choosing $1+ y_{St}\frac{v}{\Lambda}=1+ y_{Sb}\frac{v}{\Lambda}= s_t$, thus taking both Yukawas to scale as the SM Higgs Yukawa times the factor $s_t$. The partial and total width ratios of $S_{t}$ and  $S_{tb}$ compared to the SM Higgs are
\beq
\kappa^{dec}_{S_t, \gamma\gamma}\sim \kappa^{dec}_{S_{tb}, \gamma\gamma} \sim \frac{1}{8} s_t^{2}  
\ ; \quad  \frac{\Gamma[H]}{\Gamma[S_{t}]}\sim 20 s_t^{-2}  \ , \quad
\frac{\Gamma[H]}{\Gamma[S_{tb}]}\sim\frac{1}{0.6 s_t^{2}+0.05s_t^2}  \ , \eeq
which yields the total enhancement ratios:
\beq
R^{S_{t}}_{\gamma\gamma}\sim \frac{1}{8} s_t^{4} \frac{\Gamma[H]}{\Gamma[S_{t}]}  \ , \quad R^{S_{tb}}_{\gamma\gamma}\sim \frac{1}{8} s_t^{4} \frac{\Gamma[H]}{\Gamma[S_{tb}]} 
\ , \quad
\ , \quad R_{ttbb}^{S_{tb}}\sim s_t^{4} \frac{\Gamma[H]}{\Gamma[S_{tb}]}
\eeq
By construction the tree-level VBF and Higgs-strahlung production vanish here as well as the $WW^*$ and $ZZ^*$ decay modes.
We restrict $s_t $ to be bounded by $4\pi$ to maintain perturbativity of the top-Yukawa coupling and plot the $\kappa$ and $R$ factors in Fig.~\ref{Fig:bscaling}.

An important difference between $S_t$ and $S_{tb}$ is that $R_{ttbb}$ is strongly enhanced for $S_{tb}$. The current limits from Tevatron constrain $R_{ttbb}\lesssim 26$ \cite{cdftth} allowing $s_t\lesssim 4$ for $S_{tb}$. Theses are sufficiently large Yukawas leading to $R_{\gamma\gamma}^{S_{tb}}$ greater than unity. 

We estimate the signal significance for $S_t$ and $S_{tb}$ to be:
\beq
{\cal S}_{\gamma\gamma}^{S_{t}}(L)  \sim 7  \sqrt{L} \, R_{\gamma\gamma}^{S_{tb}} \ ,
{\cal S}_{\gamma\gamma}^{S_{tb}}(L)  \sim 0.64  \sqrt{L} \, R_{\gamma\gamma}^{S_{tb}} \ , \  {\cal S}_{ttbb}^{S_{tb}}(L)  \sim 0.06  \sqrt{L} \, R_{ttbb}^{S_{tb}} \  . 
\eeq
Again the estimate for ${\cal S}_{ttbb}^{S_{tb}}$ is discussed in appendix~\ref{Significance}. The estimates indicate that if e.g. $R_{\gamma\gamma}^{S_{tb}}\gtrsim 1.5$  then $S_{tb}$ would be visible in the $ttbb$ channel with the 20 ${\rm fb}^{-1}$ dataset at more than $3\sigma$. Obviously $S_t$ is not constrained by the $ttbb$ search.

\subsection{Quasi gauge phobic Scalar }
\label{sec:scalingex}
It is interesting to consider the case in which the Higgs-like  scalar $S$ couples to the SM fields like the Higgs, except for the vev entering the couplings being reduced with respect to the EW one  $v_{EW}$ by $s_v \leq 1$. This  framework is quite general and arises in several well motivated extensions of the SM such as \cite{Chivukula:2011dg}.

\begin{figure}[htp!]
\par
\begin{center}
\includegraphics[width=7.0cm]{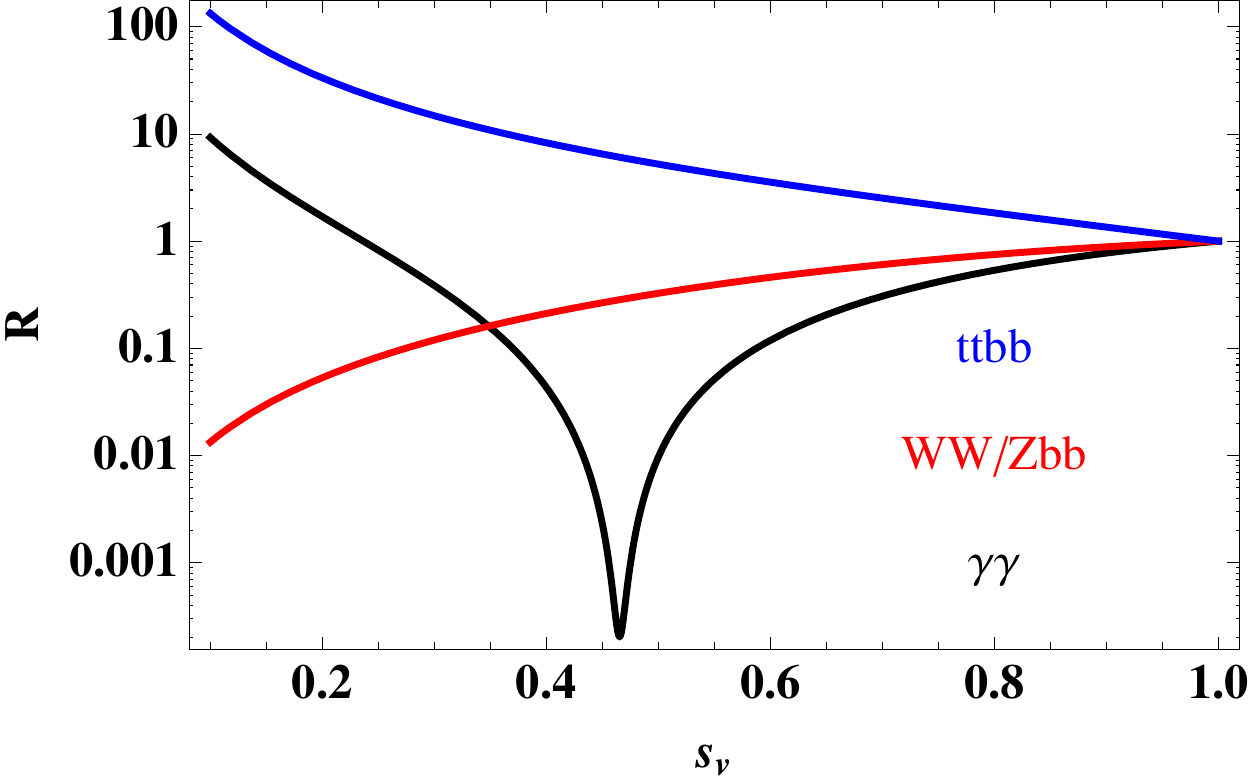}
\includegraphics[width=7.0cm]{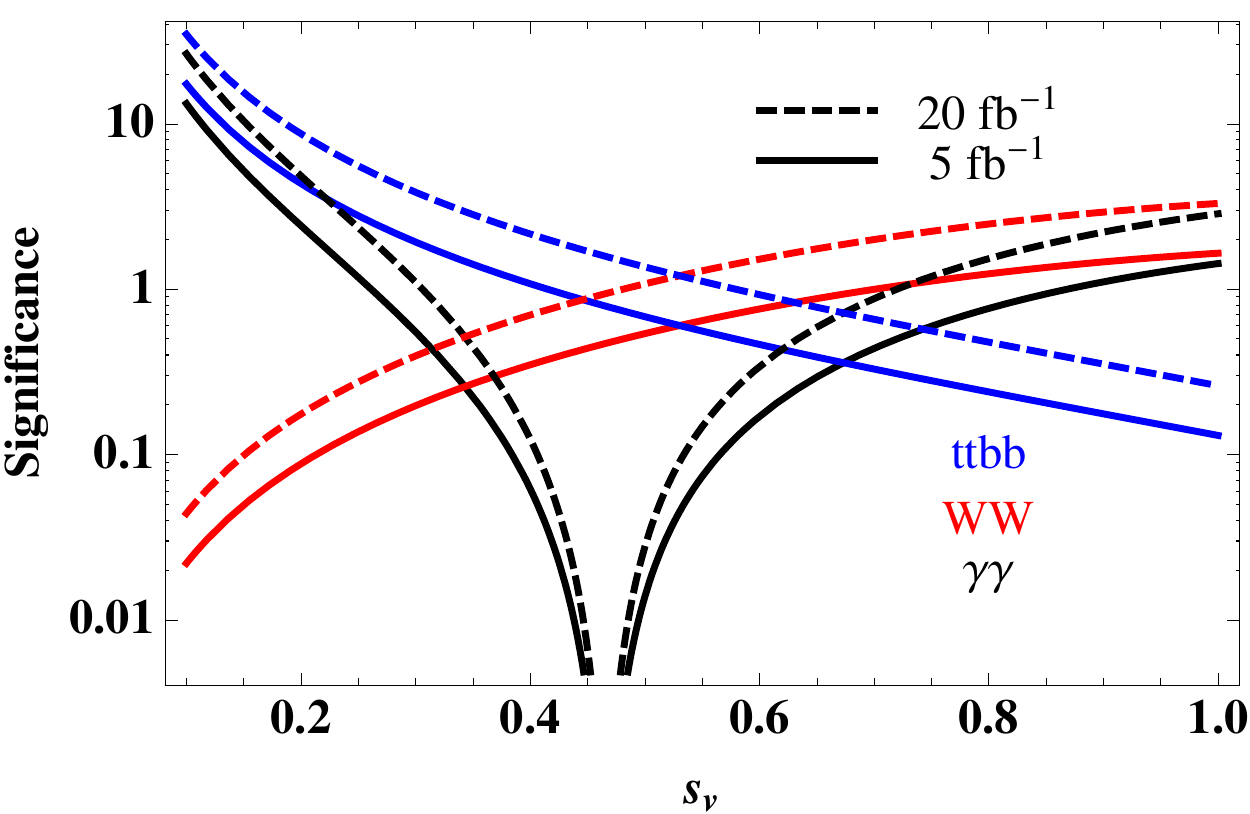}
\end{center}
\caption{Left: The enhancement ratios $R_{\gamma\gamma}$ (black), $R_{WW/ZZ}$  (red), 
$R_{ttbb}$  (blue) of $S$ as a function of the reduction $s_v$ of the electroweak scale $v$. Right: The corresponding estimates of the significance of the signals with 5 (solid) and 20 (dashed) ${\rm fb}^{-1}$ of data.
}
\par
\label{Fig:simplescaling}
\end{figure}

In this case we have $1+g_{SWW}^{(1)}\frac{\Lambda}{v}=s_v$ and the Yukawa couplings of $S$ to SM fermions $\psi$ are given by $1+ y_{S\psi}\frac{v}{\Lambda}=s_v^{-1}$. The Yukawa couplings of $S$ are thus enhanced while the massive vector boson couplings are reduced. From table~\ref{table:smbr} we then have:
\beq
\kappa^{\rm prod}_{S}&\sim& s_v^{-2} \ , \quad \kappa^{\rm prod}_{ t\bar{t} S}\sim s_v^{-2}  \ , \quad \kappa^{\rm prod}_{S V} = \kappa^{\rm prod}_{S q q',} =s_v^{2} \nonumber
\\
\kappa^{\rm dec}_{b \bar{b}} &=&  s_v^{-2}   \ , \quad  \kappa^{dec}_{\gamma\gamma }  \sim 1.8 [-s_v+\frac{1}{4} s_v^{-1}]^2 \ , \quad  \kappa^{dec}_{WW/ZZ } = s_v^2 \ , \quad
\frac{\Gamma[H]}{\Gamma[S]} \sim \frac{1}{\frac{3}{4}s_v^{-2}+\frac{1}{4}s_v^2} \ .\nonumber 
\eeq
 To arrive at the expression for $\kappa^{dec}_{\gamma\gamma }$ we have approximated the amplitude functions $A$ by their asymptotic values in Eq.~\eqref{AAA}. The corresponding $R$ ratios are given by
\beq
R_{\gamma\gamma} \sim  1.8  [s_v- 1/4 s_v^{-1}]^2 \times \frac{s_v^{-2}}{\frac{3}{4}s_v^{-2}+\frac{1}{4}s_v^2} \ , \quad 
R_{ttbb}   \sim \frac{ s_v^{-4}}{\frac{3}{4}s_v^{-2}+\frac{1}{4}s_v^2} \ , \quad 
R_{WW/ZZ} =R_{Zttbb} \sim \frac{1}{\frac{3}{4}s_v^{-2}+\frac{1}{4}s_v^2}  \nonumber \\
\eeq 
and shown in the top right panel of Fig.~\ref{Fig:simplescaling}.
Note that because we now have a non-zero coupling to both $W$ and $t$ there is a region of $s_v$ where the di-photon rate effectively vanishes.
As was the case for the $S_{tb}$ scalar, the $ttbb$ mode is significantly enhanced when reducing $s_v$. 
For example, to have an enhancement of the di-photon rate compared to the SM Higgs in this scenario we need $s_v\lesssim 0.24$, which is only just tolerated by the the current best limit from Tevatron on $R_{ ttbb}\lesssim 26$ \cite{cdftth}. This allows at most $s_v\gtrsim 0.22$ corresponding to $R_{ \gamma\gamma}$ $\lesssim 1.2$

The expected signal significance in the $ttbb$ is again correlated with that expected in the di-photon channel, ${\cal S}_{ttbb}^{S_{tb}}(L)   \sim \sqrt{L} R_{\gamma\gamma}$,
such that a di-photon cross-section greater or of the order of the SM Higgs  can be tested with circa 20 ${\rm fb}^{-1}$ in this channel.

\subsection{Pseudoscalars as natural gauge phobic states}
\label{fundamentalpseudoscalars}
\begin{figure}[htp!]
\par
\begin{center}
\includegraphics[width=6.0cm]{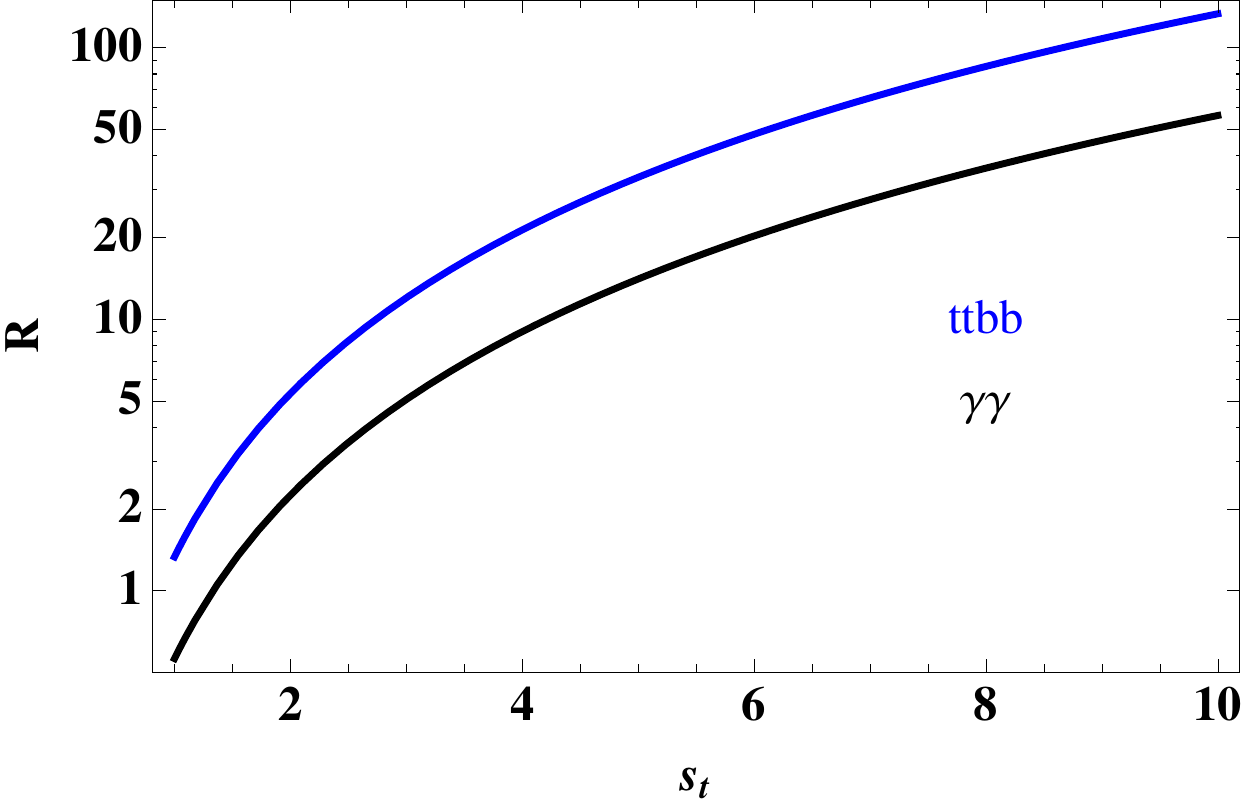}
\includegraphics[width=6.0cm]{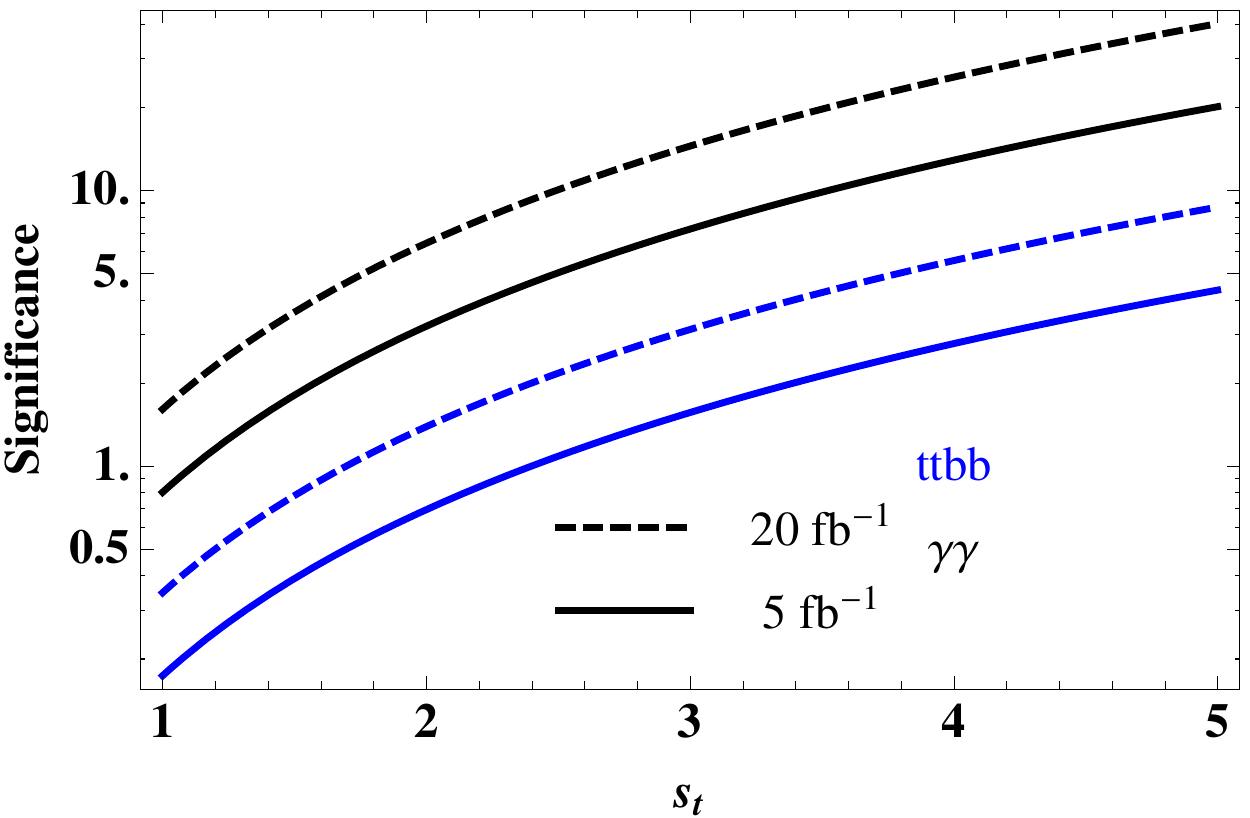}
\end{center}
\caption{Left: The enhancement factors $R_{\gamma\gamma}$ (black), and $R_{ttbb}$ (blue) for the pseudoscalar $P$ Right: The corresponding estimates of the signal significance with 5 (solid) and 20 (dashed) ${\rm fb}^{-1}$ of data.}
\par
\label{Fig:PvsSsimplescaling}
\end{figure}

A CP-odd pseudo-scalar state $P$ at a mass of $\sim 126$ GeV is generically gauge phobic --- at the dimension three operator level because we are assuming $CP$ invariance. The induced dimension five operators coupling $P$ to the SM gauge bosons arise only due to SM fermion loops, and the branching into $WW^*$ and $ZZ^*$ final states are suppressed with respect to the $\gamma\gamma$ because of phase space.  Thus comparing the widths into di-bosons of $P$ at $m_P \sim 126 \ {\rm GeV}$ and the SM Higgs with the same mass we have :
\be
\frac{\Gamma[P\to WW^*]}{\Gamma[P\to \gamma\gamma]}  \ll 1 \ , ~ \frac{\Gamma[P\to ZZ^*]}{\Gamma[P\to \gamma\gamma]}  \ll 10^{-1}
~~ {\rm vs} ~~ 
\frac{\Gamma[H\to WW^*]}{\Gamma[H\to \gamma\gamma]} \sim 10^2 \ , ~ \frac{\Gamma[H\to ZZ^*]}{\Gamma[H\to \gamma\gamma]} \sim 10  \nonumber \ .\\
\label{Eq:P}
\ee
To compare with the gauge phobic scalar $S$ case we re-express the Yukawas for $P$  such that $m_\psi \frac{y_{P \psi}}{ \Lambda} \equiv \frac{m_\psi}{v}s_t $ and arrive to the following production ratios $\kappa^{\rm prod}$:
\begin{align}
\kappa^{\rm Prod}_{P}  \sim \frac{9}{4}s_t^2 \ , \quad 
\kappa^{\rm Prod}_{P t\bar{t}} \sim  s_t^2 \ , \quad 
\kappa^{\rm Prod}_{P V_{1,2}}  \sim 0 \ , \quad 
\kappa^{\rm Prod}_{Pqq'}  \sim 0 \ ,
\end{align}
with a characteristic $9/4$ enhancement of the production via the top-loop for the gluon fusion.
The two enhancement ratios $R_{\gamma\gamma}, R_{ttbb}$ are given in the left panel of Fig.~\ref{Fig:PvsSsimplescaling} showing that  $R_{ttbb}/R_{\gamma\gamma}\sim 2.4 $.
One can see from the figure that for $s_v \lesssim 0.75$  we have that $R_{\gamma\gamma}$ is larger than unity. This can happen since  CDF limits on $R_{ttbb}$ requires only that  $s_v \gtrsim 0.23$. The strong correlation between these two channels in this scenario can be used to help unveiling the pseudo-scalar nature of $P$.  
At the LHC with 7 or 8 TeV we have ${\cal S}_{ttbb}^{S_{tb}}(L)   \gtrsim \frac{\sqrt{L}}{10} R_{\gamma\gamma}$.
To summarize this section we have discussed simple ways of testing elementary pseudo-scalars versus Higgs-like scalars at the LHC by comparing $R_{\gamma\gamma}$, $R_{WW/ZZ}$ and $R_{ttbb}$ channels. Ultimately of course, a direct CP measurement is needed to distinguish a pseudo scalar from a scalar. This could be done, for example,  using the $\tau \tau$ decay mode  \cite{ATLAS-CONF-2011-132}.

\section{Light Composite spin-0 states and Their Signatures}
\label{LCS}
We now discuss the additional effects that arise if the generic spin-0 state $\Phi$ is composite. There are two relevant new features:
First, the underlying constituent fermions will contribute to the di-photon decay rate and, if they carry ordinary color, also to the di-gluon rate. Secondly, the presence of new vector resonances can enhance the production of $\Phi$ in association with the SM gauge bosons $W/Z$. Note also that the presence of such vector resonances, mixing with the $W$ and $Z$ bosons, may also reduce the couplings of a composite scalar $S$ to $W/Z$ mass eigenstates \cite{Belyaev:2008yj}, while a composite $P$ remains essentially gauge phobic at low masses.

\subsection{Composite pions from new strong dynamics}
We first classify composite models of Technicolor according to the representation $R_\Psi$ of the new strongly interacting fermions $\Psi$. The quantum global symmetry group $G$ and its stability group $H$ (assumed to be the stability group) depend solely on the reality properties of the fermion representation as follows: 
\begin{align}
SU(N_f)_L\times SU(N_f)_R\times U(1)_V\ & \rightarrow \ SU(N_f)\times U(1)_V \quad R_\Psi \ \rm complex \nonumber
\\
SU(2N_f) \ & \rightarrow \ SO(2N_f) \quad \ R_\Psi \ \rm real \nonumber
\\
SU(2N_f) \ & \rightarrow \ Sp(2N_f) \quad  R_\Psi \ \rm pseudo-real
\end{align}
The number of massless degrees of freedom is  $N_f^2-1$ for $R_\Psi$ complex,
$2N_f^2+N_f-1$ for $R_\Psi$ real
and $2N_f^2-N_f-1$ for  $R_\Psi$ pseudo-real. It is well know that there is also a pseudoscalar, the equivalent of the QCD $\eta^{\prime}$, which is massive due to the $U(1)$ anomaly. 

The minimal Technicolor fermionic content is one weak doublet, i.e. $N_f=2$. If $R_{\Psi}$ is complex there are only the 3 NGBs becoming the $W_L$ and $Z_L$ modes and therefore there are no physical pNGBs left. If the representation is (pseudo)-real there is (one isosinglet scalar) one isotriplet pseudo-scalar pNGB's remaining in the spectrum. However, as discussed in \cite{Foadi:2007ue,Hapola:2012wi} the isotriplet pNGBs do not have Yukawa-like interactions with the SM fermions (due to the allowed hypercharge assignment) and therefore cannot be produced via gluon fusion. 
The situation changes when considering a larger number of technifermions, as in early Technicolor models \cite{Farhi:1979zx,Lane:1989ej,Appelquist:1993gi}. 
Here, the underlying technifermions also increase the di-photon decay rate compared to a fundamental pseudo-scalar. 

\subsection{Technipions in Minimal Technicolor Models}
We consider in this section technipions from minimal Technicolor theories where the underlying constituents do not carry ordinary color. These models are less constrained by both direct searches and precision measurements \cite{Sannino:2010ca,DiChiara:2010xb}. The gluon fusion production of $P$ is determined only by the size of the top-Yukawa $m_t \frac{y_{P t}}{ \Lambda} \equiv \frac{m_t}{v}s_t $ coupling while the di-photon decay width depends on both the $t$-Yukawa and the contribution from the technifermions.
For deducing the technipion  $g_{P\Psi}=y_{P\Psi} \frac{v}{\Lambda}$ in table~\ref{loopcouplingsI},  we identify $\Lambda$ with $F_\Pi$ and furthermore we take consistently $F_\Pi =\frac{v}{\sqrt{N_D}}$.  We also have that $y_{P\Psi}$ is given by the isospin charge of $\Psi$.

\begin{figure}[htp!]
\par
\begin{center}
\includegraphics[width=7.0cm]{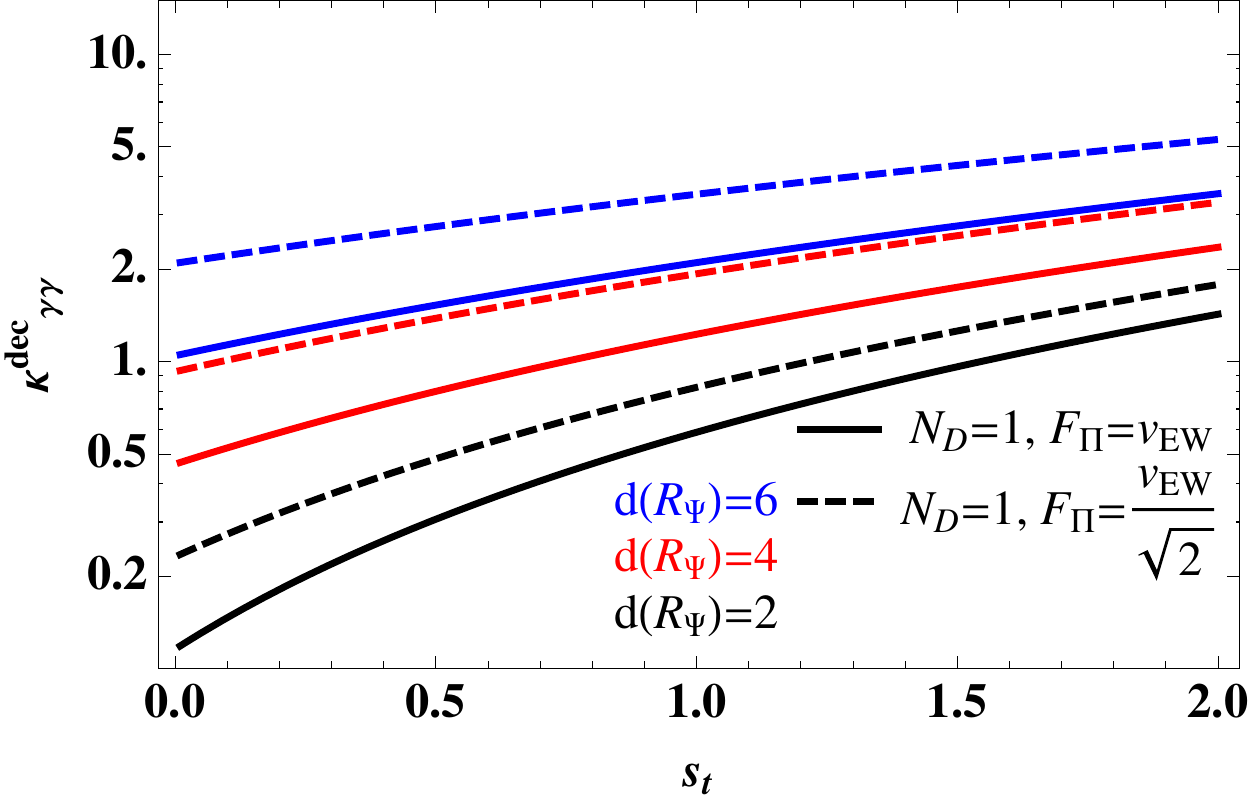}
\includegraphics[width=7.0cm]{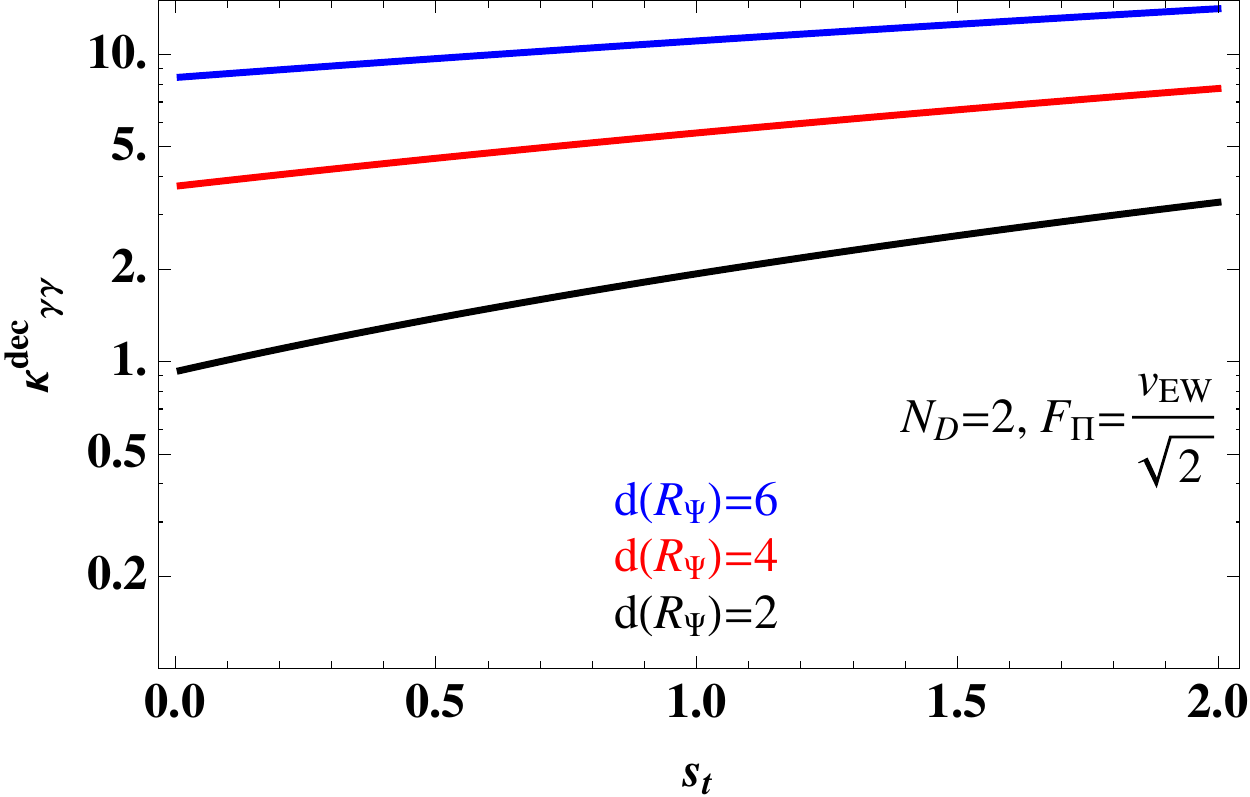}
\includegraphics[width=7.0cm]{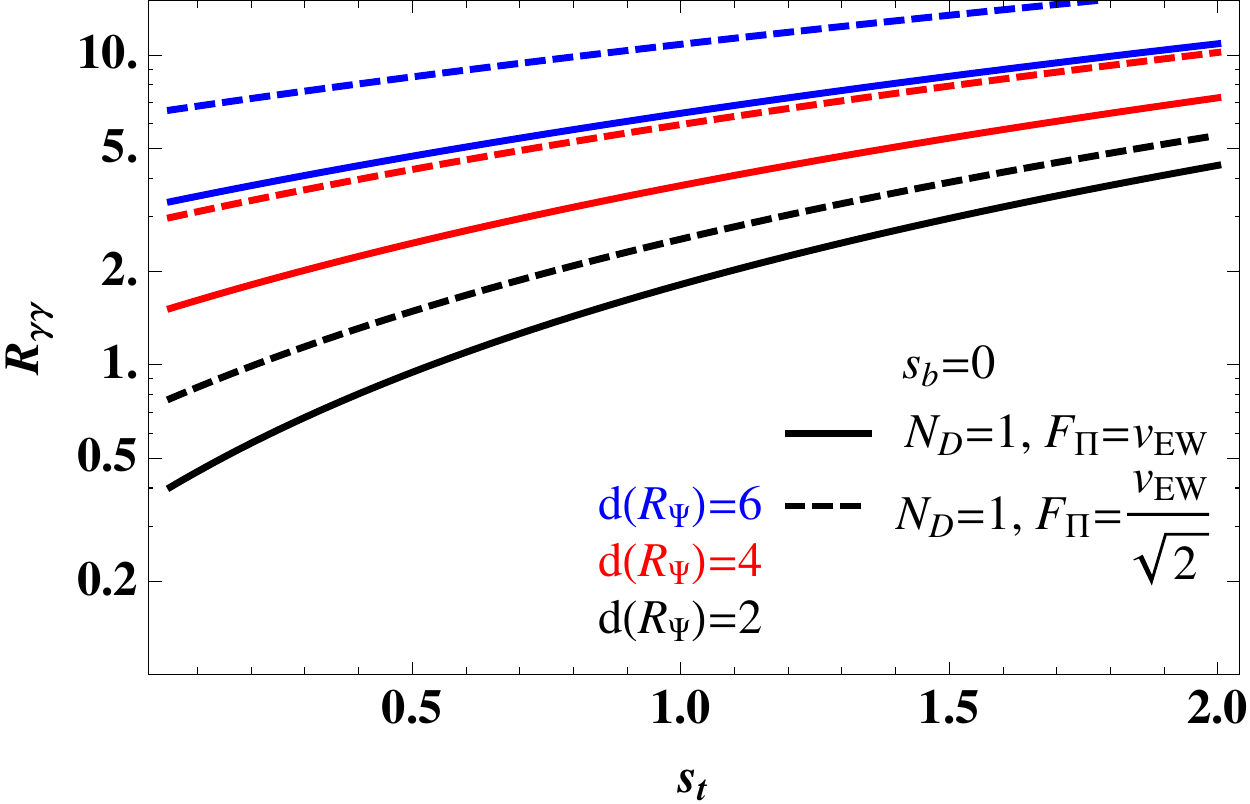}
\includegraphics[width=7.0cm]{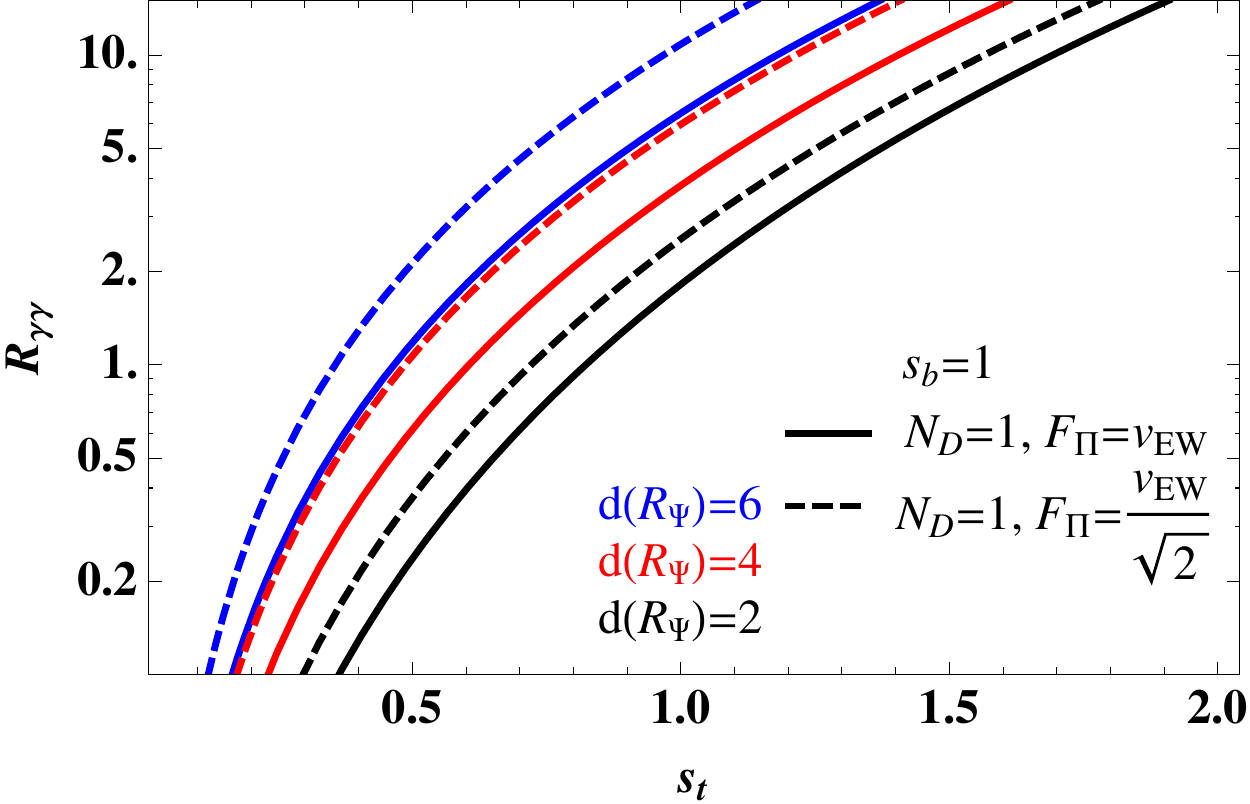}
\includegraphics[width=7.0cm]{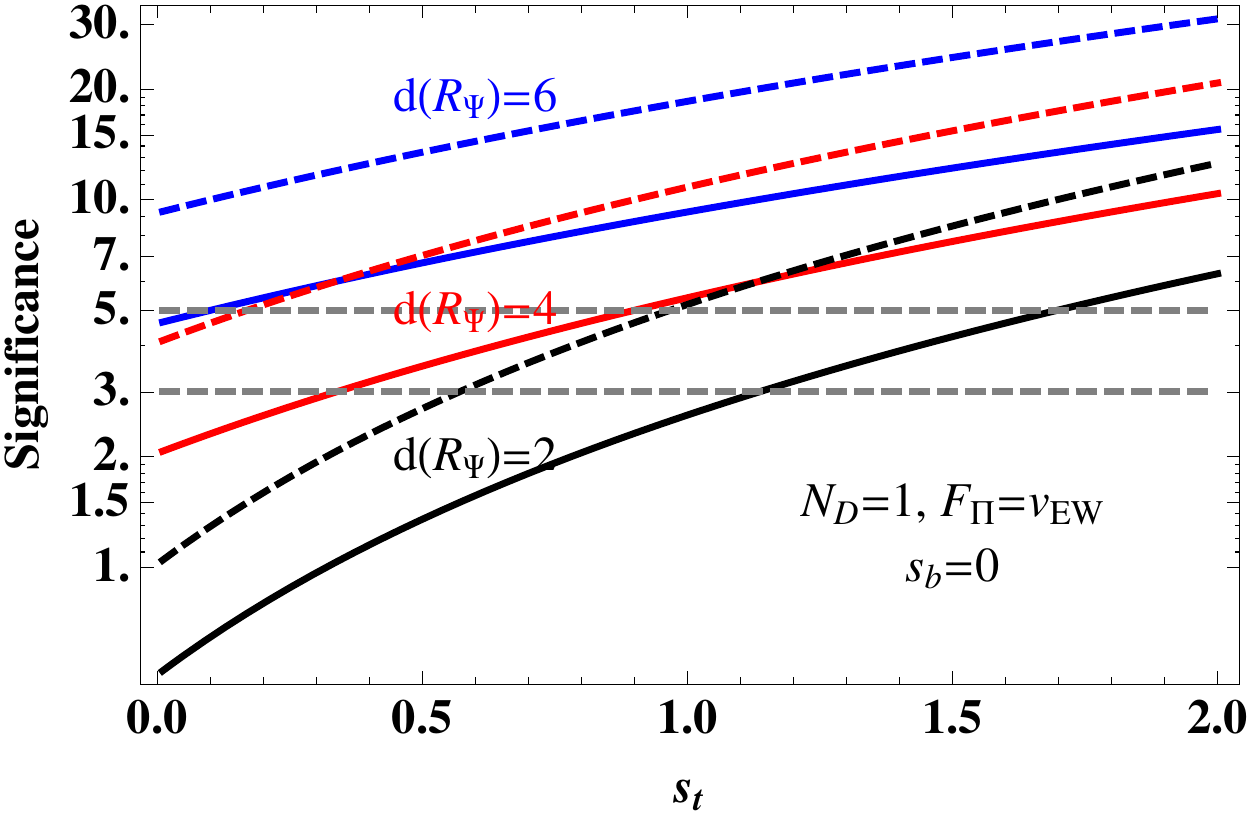}
\includegraphics[width=7.0cm]{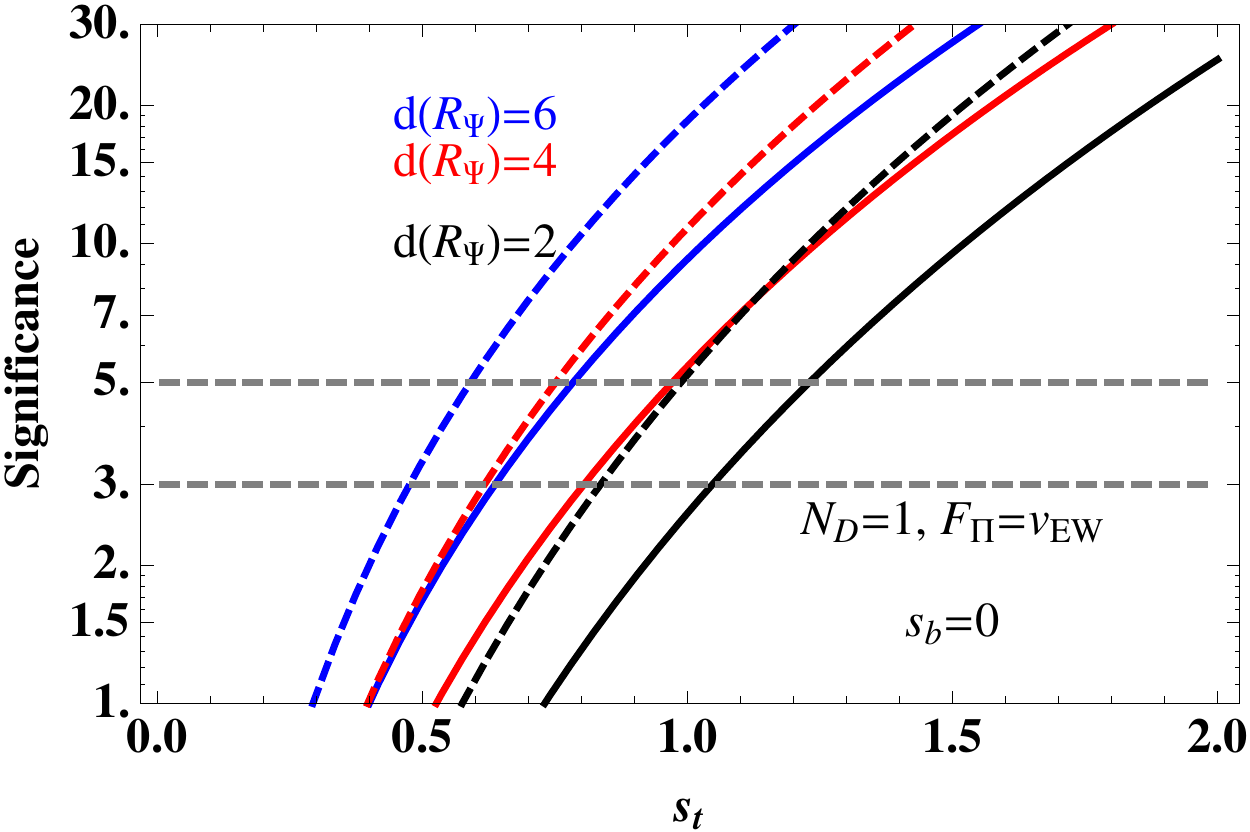}
\end{center}
\caption{Top left: The ratio  $\kappa^{dec}_{\gamma\gamma}$ of the two-photon decay width of a technipion  compared to the SM Higgs as a function of $s_t$ --- the fraction of the SM $t$-Yukawa ---  for $d(R_\Psi)=2,4,6$ (black, red, blue) and for two choices of the relation between the weak scale and $F_\Pi$ (solid and dashed lines). 
Top right: The same as top left but for $N_D=2$ such that at most $F_\Pi=v_{\rm EW}/\sqrt{2}$. 
Middle left: The corresponding ratio $R_{\gamma\gamma}$ of two-photon cross-sections for the same parameters as in the top left panel with $s_b=0$ --- the fraction of the SM $b$-Yukawa. Middle right: The same as left but with $s_b=1$.
Bottom left: The corresponding significance of the di-photon signal at LHC with 5 (solid) and 20 (dashed) ${\rm fb}^{-1}$ of data. Bottom right: The same as left but with $s_b=1$.}
\par
\label{tcPphotons}
\end{figure}

Results for ${\kappa}_{\gamma \gamma}$, for the case of an iso-singlet technipion, are given in the upper left and right panels of Fig.~\ref{tcPphotons} and the corresponding $R_{\gamma \gamma}$ factors in the middle panels with zero b-Yukawa coupling.
Even for small $s_t$, especially with more than one doublet, there is a a significant enhancement of $R_{\gamma \gamma}$. Of course this enhancement is reduced when turning on a $b$-Yukawa for  $P$ as shown in the third row of Fig.~\ref{tcPphotons}.
As it is clear from the figures, the $t$-Yukawa of the composite $P$ to SM fermions must typically be smaller than one in order not to be ruled out by current di-photon data, even at a mass $m_P\sim 126$ GeV corresponding to the observed excess.
The difference with the elementary pseudo-scalar case is that there the di-photon cross-section is determined solely by the $t$-Yukawa coupling --- up to the width effects of the $b$-Yukawa. Therefore the $ttbb$ search channel may be used to disentangle a composite pseudo-scalar from an elementary one if appearing in di-photon searches. 

Another interesting framework yielding a composite pseudoscalar with $t$-Yukawa couplings larger than the SM Higgs is top-color \cite{Hill:1994hp}. The neutral top-pion is thus a certain realization of the pseudoscalar in section ~\ref{fundamentalpseudoscalars} and were recently considered in detail in \cite{Chivukula:2011ag}. 
Finally, if the technifermions carry ordinary color as studied in e.g.  \cite{Lubicz:1995xi,Belyaev:2005ct,Bernreuther:2010uw,Chivukula:2011ue} it is easy to understand that they have larger enhancements for both production and decay. For these states the current LHC data is clearly imposing strong constraints.

 \subsection{Composite Higgs}
\begin{figure}[htp!]
\par
\begin{center}
\includegraphics[width=7.0cm]{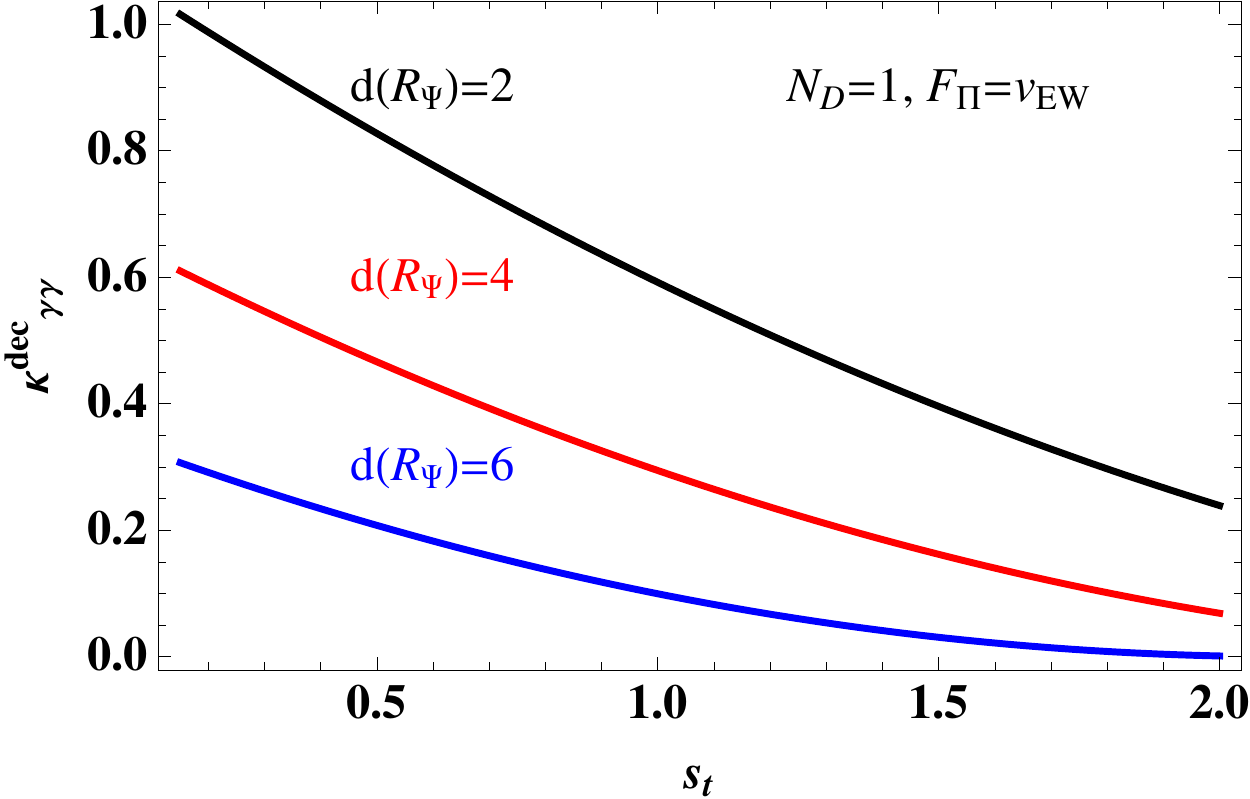}
\includegraphics[width=7.0cm]{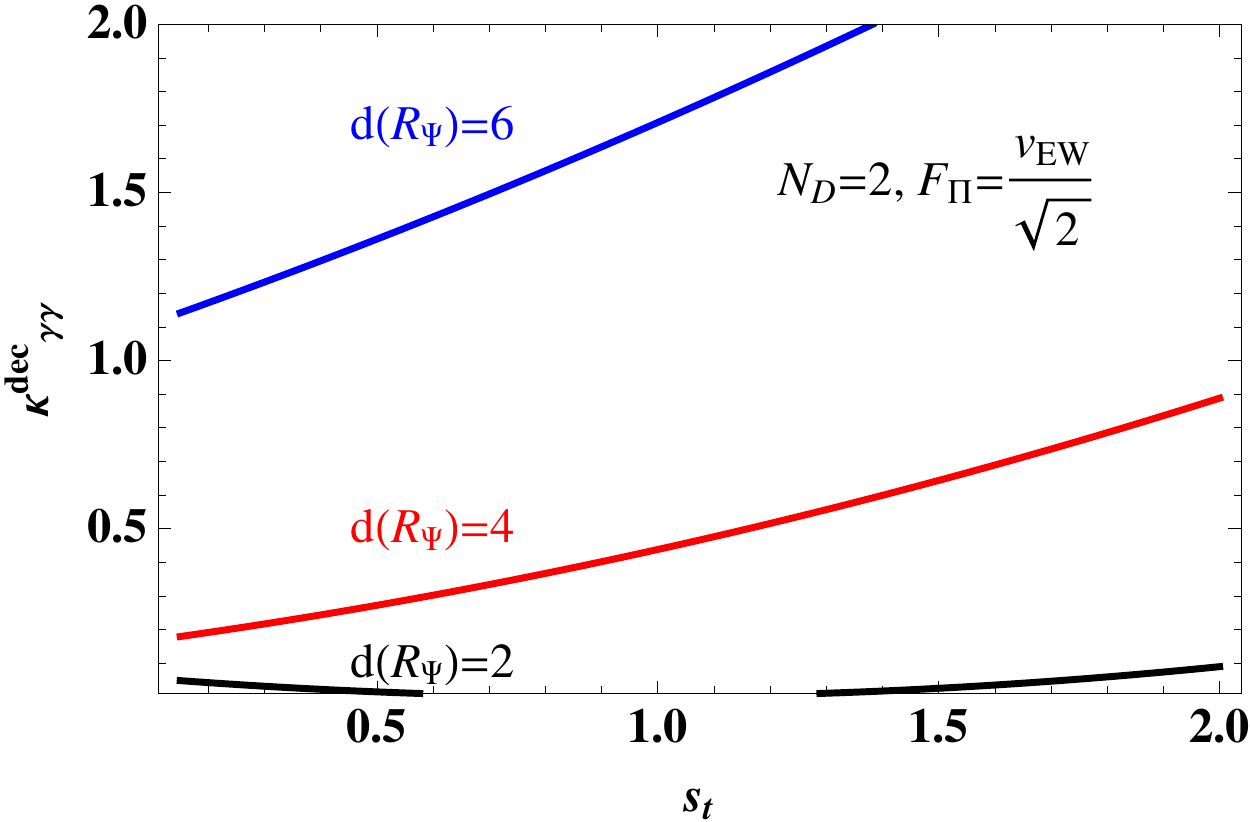}
\includegraphics[width=7.0cm]{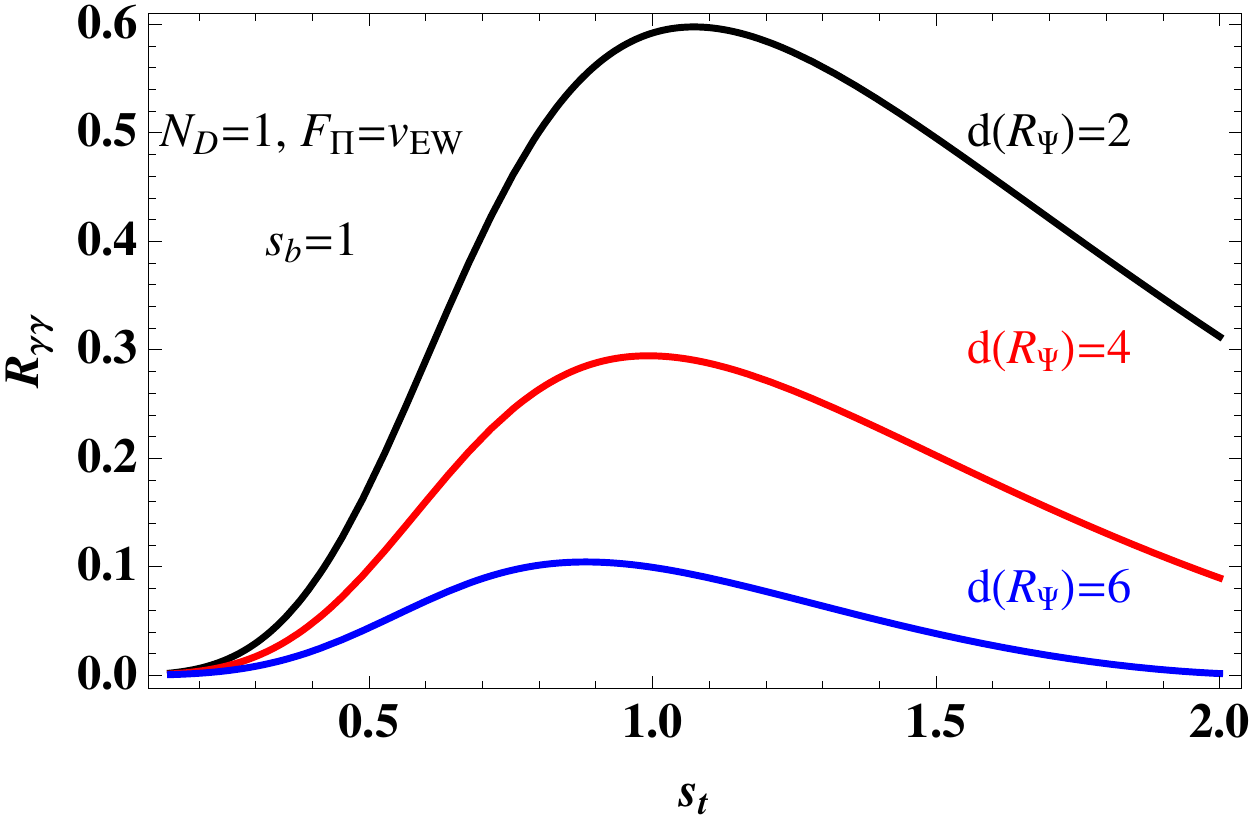}
\includegraphics[width=7.0cm]{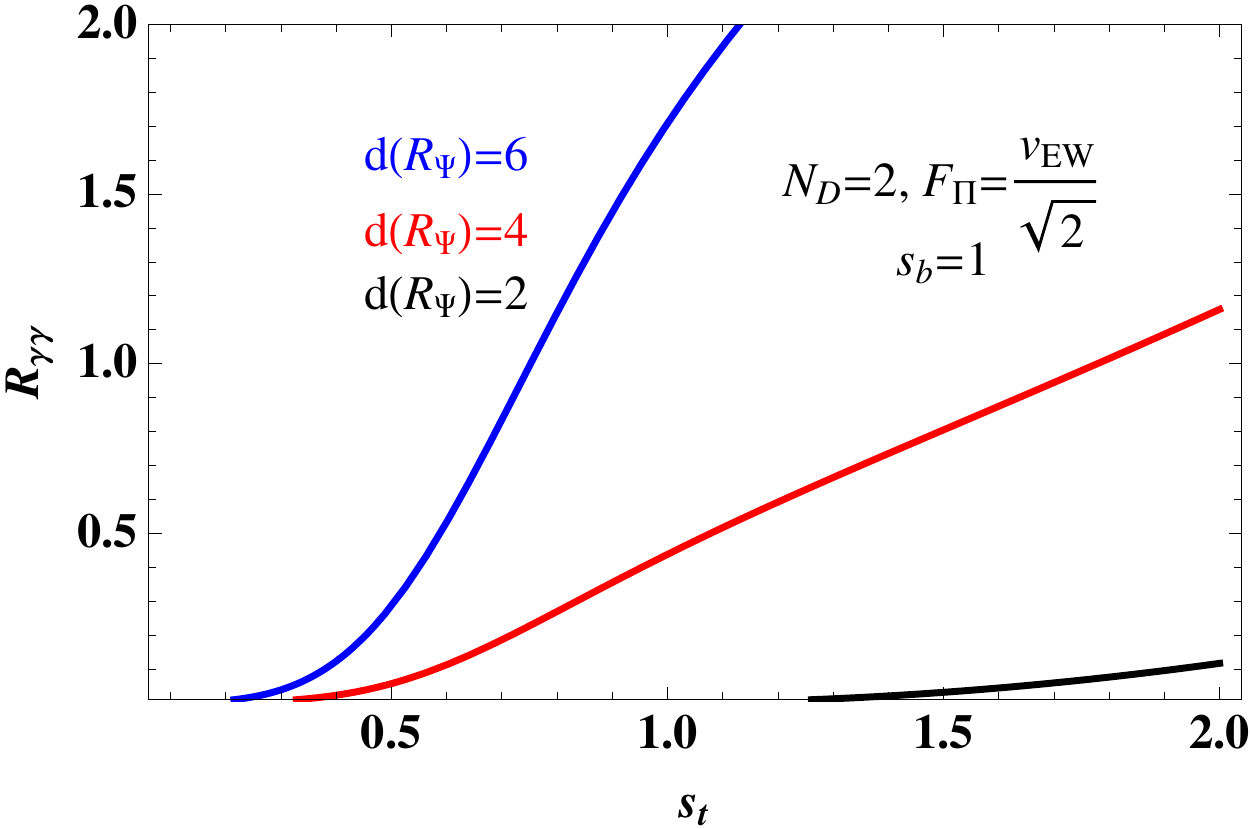}
\end{center}
\caption{Top left: The ratio $\kappa_{\gamma\gamma}^{\rm dec}$ of two-photon decay widths of a composite scalar $S$ compared to the SM Higgs as a function of $s_t$, the fraction of the SM $t$-Yukawa, and with $d(R_\Psi)=2,4,6$ (black,red,blue). Top Right: The same as left but with $N_D=2$ and thus $v_{\rm EW}=F_\Pi/\sqrt{2}$.
Bottom left: The corresponding ratio $R_{\gamma\gamma}$ of two-photon cross-sections for the same parameters as in the top left panel. 
Bottom right: The same plot as left but with $N_D=2$ and thus $v_{\rm EW}=F_\Pi/\sqrt{2}$: 
}
\par
\label{Fig:TChiggs}
\end{figure}

The di-photon decay width of a light composite Higgs \cite{Dietrich:2005jn} or {\it techni-dilaton} \cite{Yamawaki:1985zg} has been studied in e.g. \cite{Hapola:2011sd,Matsuzaki:2012gd}. Again we first consider models with colorless technifermions. 
The difference between a tecnipion and a composite Higgs is that for the composite Higgs, the contribution from technifermions interferes destructively with the otherwise dominant effect of the SM W-loop. In order to get a large branching ratio into di-photons we either need a large number of new technifermions or a low scale in the Technicolor sector.  These features are clear from Fig~\ref{Fig:TChiggs} where we show $\kappa_{\gamma\gamma}^{\rm dec}$ and $R_{\gamma\gamma}^{\rm dec}$ for a composite Higgs with different values of $R_\Psi, N_D$ as well as the top Yukawa coupling.

\subsection{Resonantly enhanced associate production of a composite spin-0 state}
We finally study resonant enhancement of the production cross-section of a composite spin-0 state in association with a SM $W$ or $Z$ gauge boson, via a new spin one composite state $R$. This possibility was discussed for the composite Higgs case in \cite{Zerwekh:2005wh,Belyaev:2008yj} and for technipions in a different context in e.g. \cite{Eichten:2011sh}. In principle, the cross section for the production of a generic (pseudo)scalar field in association with a $W$ or a $Z$ boson can be enhanced to the level of the gluon fusion production of the SM Higgs. If the composite state is near the fermiophobic limit then even a minimal Technicolor  model can give a sizable di-photon cross-section.

\begin{figure}[htp!]
\par
\begin{center}
\includegraphics[width=7.0cm]{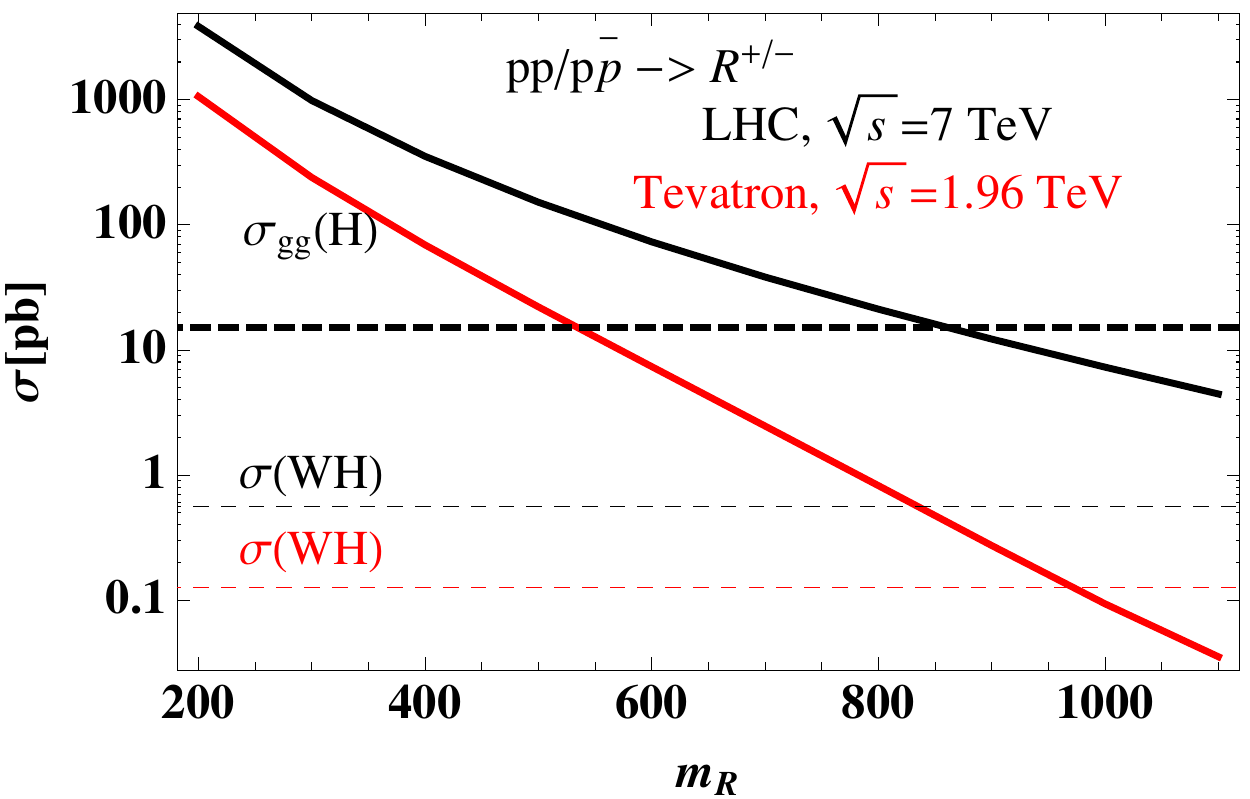}
\includegraphics[width=7.0cm]{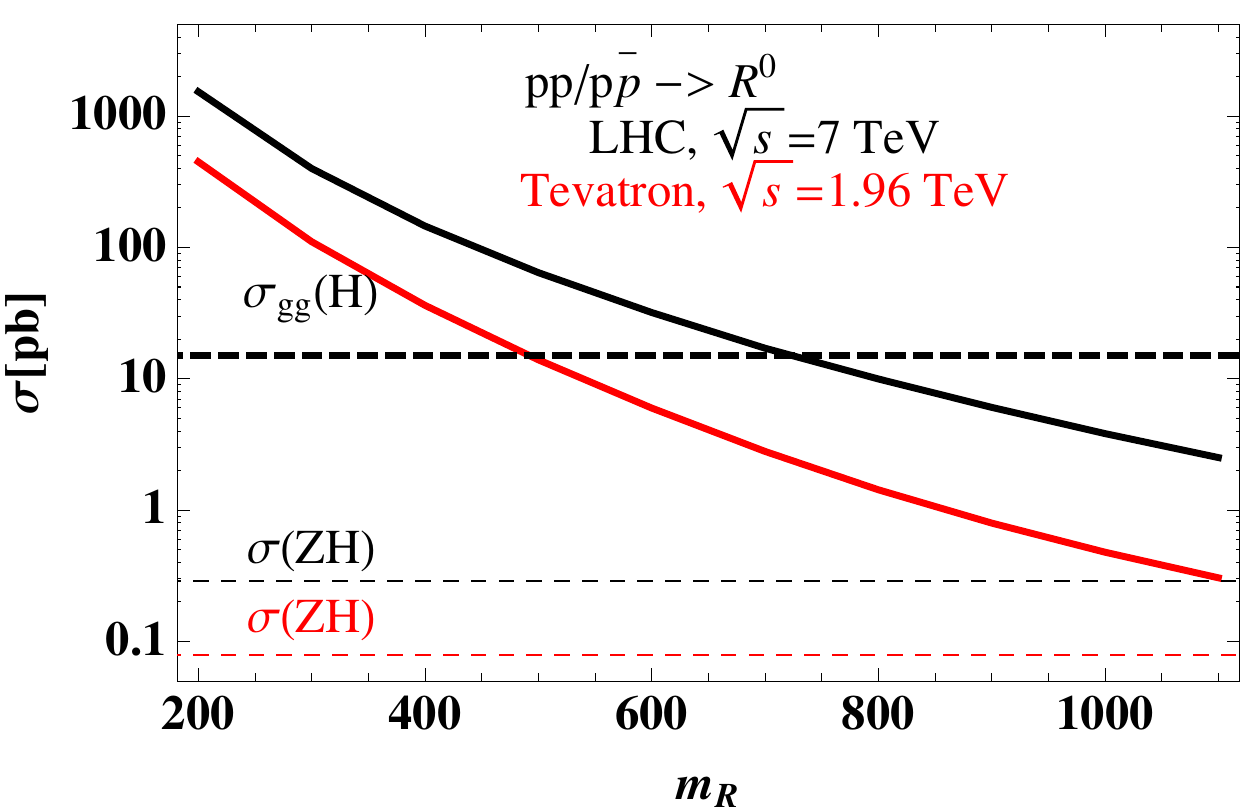}
\end{center}
\caption{Left: Drell-Yan production cross-section of $R^\pm$ as a function of mass at LHC (black) and the Tevatron (red) with fermion couplings equal to the SM $W^\pm$ bosons. Right: The same for the neutral resonance $R^0$ with fermion couplings equal to the SM $Z$ boson.
}
\par
\label{Fig:associateprod}
\end{figure}

We denote the neutral vector resonance by $R^0$ and the charged ones by $R^\pm$ \footnote{Here we are considering either a (mostly) axial or vector massive spin-one state to appropriately match the relevant quantum numbers of either the composite Higgs or pseudoscalar \cite{Belyaev:2008yj}}. 
We are interested in the cross-section
\beq
\label{Rgammagamma}
\sigma(pp\to R \to Z \gamma\gamma ) \simeq \sigma(pp\to R) 
{\rm BR}[R \to Z \Phi]{\rm BR}[\Phi \to \gamma\gamma] \ .
\eeq
A similar contribution to the associate production with $W^{\pm}$ can arise from $R^\pm$. 
It is straightforward to describe the relevant interactions of $R$ via the effective Lagrangian 
\beq
\mathcal{L}=\sum_{\psi=u,d} R_\mu \, \bar{\psi}\gamma^\mu(g_{V \psi}^R + g_{A\psi}^R \gamma^5)\psi + g^2 \frac{\Lambda}{2} g_{\Phi R Z} R_\mu Z^\mu \Phi  + \nonumber 
\\
R_\mu^+ \, \bar{u}\gamma^\mu g_{L\, ud}^R P_L d + g^2 \frac{\Lambda}{2} g_{\Phi R W} R_\mu^+ W^{\mu -} \Phi + h.c 
\ . 
\eeq
In figure~\ref{Fig:associateprod} we provide the production cross-section of the $R^{0,\pm}$ resonances. We choose the values of the couplings $g^{R}_{V,A\psi}$ equal to the equivalent ones for the SM $W$ and $Z$ bosons \footnote{These couplings are labeled by $g_{A,V \, u,d}^{Z}$ and for the SM $Z$ boson assume the following values $g_{V u}^{Z}=\frac{e}{4 s_W c_W}(1-\frac{8}{3}s_W^2)\sim 0.07$, $g_{V d}^{Z}=- \frac{e}{4 s_W c_W}(1-\frac{4}{3}s_W^2)\sim -0.13$, $g_{A u}^{Z}=-\frac{e}{4 s_W c_W}\sim -0.19$, $g_{A d}^{Z}=-\frac{e}{4 s_W c_W}\sim 0.19$.}. This choice corresponds to the values of the couplings of a sequential $W'$ and $Z'$ model.
 
If ${\rm BR}[R^{0,\pm} \to Z/ W^\pm \,  \Phi]\sim 1$ the cross-section for associate production of $ Z/ W^\pm \,  \Phi $ is identical to the DY production cross-section of $R^{0,\pm}$. We therefore also show, for comparison, the gluon-fusion and associate production of the SM Higgs at 125 GeV in the figures. The Tevatron cross-section has been taken from \cite{Baglio:2010um}.
Obviously our normalization choice of the $R$ couplings to the SM fermions can be fitted to any other model by a simple rescaling. For example, in the (Next to) Minimal Walking Technicolor models, the axial vector resonance does have a branching ${\rm BR}[R \to W \Phi]\sim 1$ to the composite Higgs while the coupling $g^{R}$ is  $\lesssim 0.3 g^Z$  \cite{Belyaev:2008yj}.  
This value of still allows for an enhancement of the associate Higgs production for $m_R \sim 500-600$ GeV \footnote{A sequential SM Z' is ruled out at such masses. However the resonances considered here are not ruled out because the coupling to SM fermions is suppressed \cite{Belyaev:2008yj}.} making it comparable the gluon fusion production of the SM Higgs.

 \section{Summary}
 \label{summary}
 Using an effective Lagrangian approach  we investigated the ability to discriminate, with current and upcoming experimental data, between a scalar and a pseudoscalar stemming from several extensions of the standard model.  We also discussed how to disentangle whether such a new spin-0 state is elementary or composite. 
 
We first investigated an elementary spin-0 state, in the limit where it is fermiophobic or gauge phobic. We also investigated some intermediate cases such as a $b$-phobic state. 

The pseudoscalar, in the investigated mass energy range,  is naturally gauge phobic and therefore should not lead to any $WW^\ast$ and $ZZ^\ast$ signals with respect to the scalar case. The elementary pseudoscalar di-photon signal can be enhanced with respect to the SM Higgs if the top-Yukawa like coupling is also enhanced. This could potentially be tested in the $ttbb$ final state. 

For the scalar case the situation is more involved because of the interplay between the Yukawa and gauge boson couplings. Of course, in the gauge phobic limit the scalar has properties similar to the pseudoscalar. As already shown in the literature a $b$-phobic scalar also enhances the di-photon signal. We pointed out however, that within this years data taking one can confirm the existence of a $b$-phobic scalar by studying the $WW^\ast$ and $ZZ^\ast$ final states which should reach a significance level of $5\sigma$.

In the second part of the paper we investigated signals from composite scalars and pseudscalars stemming from extensions of the SM of Technicolor type, in which the underlying technifermions do not carry ordinary color. Several features are similar to the elementary case depending on the specific Technicolor extension. However, due to the presence of new technifermions, the composite pseudoscalar decay rate into di-photons is enhanced and this channel can ultimately be used to disentangle the underlying Technicolor dynamics. 

Finally we explored production of the composite spin-0 state in association with a SM gauge boson. This process can be enhanced due to the presence of new composite vector bosons. 
In fact, with a light composite vector resonance  the associate production can become comparable to or even exceed the SM Higgs gluon-fusion production. This leads to the interesting possibility of a sizeable di-photon signal, comparable to the SM Higgs, even for a fermiophobic composite scalar. 

We provided, for the various proposed signals, the relevant significance estimates for $5$ and $20~{\rm fb}^{-1}$ corresponding to current and expected LHC integrated luminosity for 2012. 
 
\acknowledgements
We thank G. Azuelos, A. Barr, G. Choudalakis, J. Ferrando, F. Kahlhoefer, C. Hays, and K. Schmidt-Hoberg for comments, discussions and suggestions. 
 
\appendix

\section{Significance estimate for the $ttbb$ channel}
\label{Significance}

We can make a rough estimate of the signal significance at LHC in the $ttbb$  channel with luminosity $L$, particularly interesting for the scalar $S_{tb}$.  
In \cite{Plehn:2009rk}, utilizing top-tagging techniques, a significance ranging from 4.5 to 2.9 $\sigma$ for the SM Higgs in the mass range 120-130 GeV was found at the LHC, with 100 ${\rm fb}^{-1}$ at 14 TeV. 
We do a rough extrapolation by  assuming the fraction of scalars $S_{tb}$ with $p_T(S_{tb})>200$ GeV is the same at 14 TeV and 7 TeV (we find this to be reasonably correct at parton-level). Then we take the ratio of the production cross-section in the top-fusion channel between 14 and 7 TeV to be a factor of $6$ from \cite{ATLAS:1278455}.  
 
Without scaling the backgrounds used in \cite{Plehn:2009rk}
and normalizing to a significance of $3.5$ for a 126 GeV SM Higgs with 100 ${\rm fb}^{-1}$ at 14 TeV we then have
\beq
{\cal S}_{ttbb}^{S_{tb}}(L) \sim \frac{3.5}{6} \sqrt{\frac{L}{100}} R_{ttbb}^{S_{tb}} \sim 0.06  \sqrt{L} \, R_{ttbb}^{S_{tb}} \ .
\eeq
While these estimates should be made more precise it is interesting to note that if e.g. $R_{\gamma\gamma}^{S_{tb}}\gtrsim 1.5$ as consistent with the current di-photon then $S_{tb}$ would be visible in this channel with the 20 ${\rm fb}^{-1}$ dataset at more than $3\sigma$.

\bibliography{pigamma}
\bibliographystyle{ArXiv}

\end{document}